\documentclass[twocolumn,english,aps,superscriptaddress,nolinenumbers,nofootinbib]{revtex4-1}
\usepackage[T1]{fontenc}
\usepackage[latin9]{inputenc}
\usepackage{amsmath}
\usepackage{amssymb}
\usepackage{graphicx}

\makeatletter
\usepackage{babel}

\usepackage{babel}

\usepackage{babel}

\makeatother

\usepackage{babel}
\begin{document}

\title{Vectorial dissipative solitons in vertical-cavity surface-emitting
lasers with delays}

\author{M. Marconi}

\affiliation{Institut Non Linéaire de Nice, Université de Nice Sophia Antipolis
- Centre National de la Recherche Scientifique, 1361 route des lucioles,
F-06560 Valbonne, France}

\author{J. Javaloyes}

\affiliation{Departament de Fisica, Universitat de les Illes Baleares, C/ Valldemossa,
km 7.5, E-07122 Palma de Mallorca}

\author{S. Barland}

\affiliation{Institut Non Linéaire de Nice, Université de Nice Sophia Antipolis
- Centre National de la Recherche Scientifique, 1361 route des lucioles,
F-06560 Valbonne, France}

\author{S. Balle}

\affiliation{Institut Mediterrani d'Estudis Avançats, CSIC-UIB, E-07071 Palma
de Mallorca, Spain}

\author{M. Giudici}

\affiliation{Institut Non Linéaire de Nice, Université de Nice Sophia Antipolis
- Centre National de la Recherche Scientifique, 1361 route des lucioles,
F-06560 Valbonne, France}
\begin{abstract}
We show that the nonlinear polarization dynamics of a vertical-cavity
surface-emitting laser placed into an external cavity leads to the
formation of temporal vectorial dissipative solitons. These solitons
arise as cycles in the polarization orientation, leaving the total
intensity constant. When the cavity round-trip is much longer than
their duration, several independent solitons as well as bound states
(molecules) may be hosted in the cavity. All these solutions coexist
together and with the background solution, i.e. the solution with
zero soliton. The theoretical proof of localization is given by the
analysis of the Floquet exponents. Finally, we reduce the dynamics
to a single delayed equation for the polarization orientation allowing
interpreting the vectorial solitons as polarization kinks.
\end{abstract}
\maketitle

\section{Introduction\label{introduction}}

The field of dissipative solitons (DS) in optical systems has been
the subject of an intensive research during the last twenty years
(see e.g. \cite{AA-LNP-05,AA-LNP-08,AFO-AAM-09,DC-LNP-11,GA-NAP-12}
and references therein). Optical DS are localized light pulses in
time or localized beams in space appearing in nonlinear systems kept
out of equilibrium by a continuous flow of energy counteracting the
losses. As a consequence, they exhibit remarkable differences with
respect to their conservative counterparts which arise purely as the
result of a nonlinearity compensating for the spreading effects. 

DS are \emph{attractors}, i.e. stable solutions towards which the
system evolves spontaneously from a wide set of initial conditions
\cite{NP-SelfOrg-77}. It entails that, at variance with their conservative
analogues, DS do not rely on a proper seeding of the initial conditions.
This renders them extremely interesting for applications and, in particular,
for information processing. As they coexist with the background solution,
they can be individually addressed and used as information bits \cite{CoulletLSinfo}.
All-optical buffers were demonstrated using both spatial and temporal
DS \cite{BTB-NAT-02,LCK-NAP-10,AFO-AAM-09,CSLKuszelewicz,GBG-PRL-08}. 

Another important aspect concerns the role of the nonlinearity. While
in the beginning DS were envisioned as weakly modified conservative
solitons \cite{FT-PRL-90}, it was latter shown that for strong dissipation,
the role of the nonlinearity leading to their formation is not limited
to the compensation of the spreading effect. Dissipative Solitons
are known to occur both in the normal and the abnormal dispersion
regimes \cite{CSA-PRA-09}. For instance, the so-called Cavity Solitons
appearing in the transverse plane of driven resonators \cite{rosanov,TML-PRL-94,FS-PRL-96,1172836,BTB-NAT-02,LCK-NAP-10}
have been described as \emph{localized structures}, i.e. as elementary
cellular patterns (or cellular pulses) generated by fronts connecting
different coexisting spatial solutions \cite{PhysRevLett.84.3069}.
In general, CS may appear even in presence of defocussing nonlinearities,
i.e. which \emph{favor} the spatio-temporal spreading effect \cite{1172836}. 

Temporal DS have been largely studied in long-cavity mode-locked fiber
lasers \cite{GA-NAP-12}, and several interesting behaviors have been
experimentally demonstrated, as e.g. soliton bound states \cite{TMT-PRA-01},
molecules \cite{SPM-PRL-05,GS-LNP-08}, repulsive/attracting forces
on extremely long scale \cite{TVZ-PRL-12,JEM-NAP-13}, soliton rain
\cite{CG-PRA-10} or soliton explosion \cite{CSA-PRL-02}. When the
vectorial degree of freedom of the light is taken into account like
e.g. in the case of the Manakov solitons \cite{KSA-PRL-96}, more
complex behaviors are observed. Anti-phase switching between orthogonally
linearly polarized states has been recently observed \cite{QGR-PRA-97,ZTZ-PRB-09}
and interpreted in terms of soliton domain walls, i.e. localized states
separating domains of emission in orthogonal polarizations \cite{LGW-JOSAB-13,ZTZ-PRB-09}. 

In this manuscript we report evidence of temporal vectorial DS in
a single mode Vertical-Cavity Surface-Emitting Laser (VCSEL) enclosed
into a polarization sensitive double external cavity. By exploiting
the two different times of flight as well as the polarization selectivity
of the cavity, we are able to control the polarization state of light
and generate vectorial DS. They consist in time-localized rotations
of the emitted polarization orientation, thus leaving the total intensity
constant. When the cavity round-trip is much longer than the duration
of the vectorial DS, several independent DS and/or DS bound states
(molecules) may be hosted in the cavity where they coexist with the
background solution (zero DS emission). We show that molecules and
independent DS can be discriminated experimentally by analyzing their
noise-induced motion. 

The vectorial nature of the reported DS is intimately related to the
nonlinear polarization response of the VCSEL. Because of their axial
symmetry, these devices lack strong enough anisotropies to pin the
polarization orientation \cite{CRL-APL-94,CSL-JQE-95,MFB-APL-05}
and competition between orthogonal linearly polarized states is easily
induced by polarized perturbations. Such dynamics doesn't involve
a strong exchange of energy between the light and the active medium,
and the addressing speed of these DS shall not be limited by the nanosecond
timescale governing the evolution of the gain. While the complex polarization
dynamics of VCSEL was recently exploited to generate chaotic states
for data encryption \cite{VP-NAP-13}, we show in this manuscript
that it can also be harnessed for all-optical information encoding
where bits of information are stored in the form of vectorial DS as
short as 25~ps, as predicted by our analysis.

From the theoretical point of view, it is important to point out that
the nonlinearity in our system is concentrated on a single point (the
VCSEL) rather than being distributed along the propagation direction
as in fiber resonators. For this reason, we model our DS by Delay
Differential Equations (DDEs) rather than by Partial Differential
Equations (PDEs), as the prototypic cubic-quintic Ginzburg-Landau
equation of DS. In fact DDEs possess the same complexity as PDEs since
they both correspond to dynamical systems of infinite dimensionality.
Moreover, conceptual links between PDEs and DDEs do exist. It was
revealed that a delayed system close to an Andronov-Hopf bifurcation
can be described via a Ginzburg-Landau equation \cite{GP-PRL-96}
and recently, a mapping between a spatially extended laser cavity
and an ensemble of coupled Delay Algebraic Equation was developed
\cite{JB-OE-12,PJB-JSTQE-13}. In this manuscript, we establish a
criterion for temporal localization in DDE systems based upon the
degeneracy of the Floquet exponents. Finally, we reduce the full model
to a single delayed equation for the polarization orientation allowing
interpreting the vectorial solitons as polarization kinks and anti-kinks.

\section{Experimental results\label{experimental}}

\begin{figure}[ht]
\centering{}\includegraphics[width=0.4\textwidth]{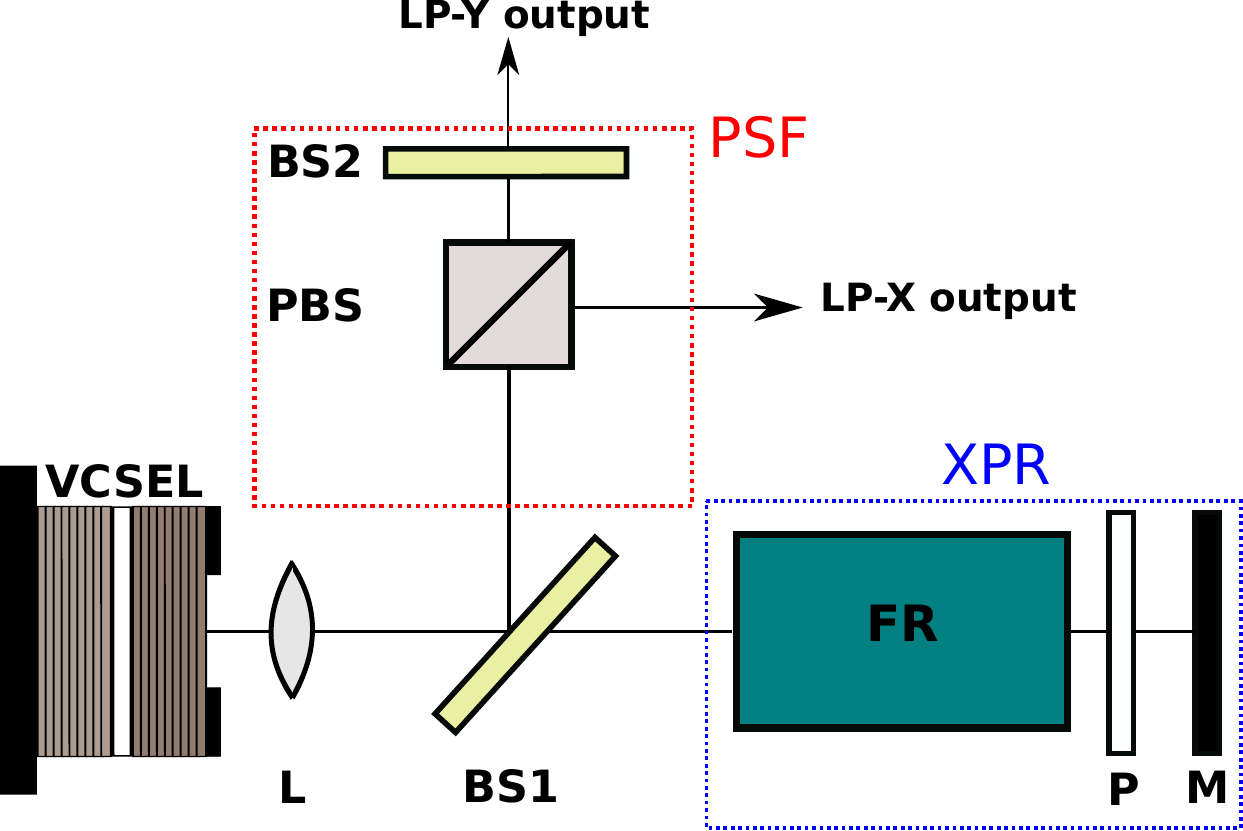}
\caption{Experimental set-up. The light emitted is collimated by an aspheric
lens (L) and split in two beams by a polarization preserving beam-splitter
(BS1). In the PSF branch, a polarizing cube (PBS) transmits only the
Y component towards a partially reflective mirror (BS2) that feeds
back Y into the VCSEL. The PBS sends the X component out of the cavity
for detection while the detection of the Y component is obtained using
the beam transmitted by BS2. In the XPR branch, the beam is sent to
a $\pi/4$ Faraday rotator (FR) with an exit polarizer (P) transmitting
only the Y component that will be rotated into X. Then, the mirror
M reflects the light back towards the VCSEL. Because of the double
passage into the FR, Y is finally re-injected into the VCSEL with
a polarization axis parallel to X. Neutral density filters in both
branches allow varying feedback levels.\label{fig:setup} }
\end{figure}

The experimental setup is schematically shown in Fig.~\ref{fig:setup}
and more details can be found in Methods~\ref{sub:Setup-and-VCSEL}.
A single-transverse mode VCSEL is coupled to a double external cavity
that selects one of the linearly polarized states of the VCSEL (Y,
say) and feeds it back twice into the VCSEL: once into the same polarization
and once into the orthogonal one. The first arm provides Polarization
Selective Feedback (PSF) with a time delay $\tau_{f}$, while the
second arm re-injects Y into the orthogonal polarization orientation
with time delay $\tau_{r}$. Such Crossed-Polarization Re-injection
(XPR) induces cross-gain modulation of the two linear polarization
components \cite{MFB-APL-05}, thus enhancing their competition and
leading to polarization dynamics \cite{MJG-PRA-13,JMG-PRA-14}. This
double cavity was shown to promote Passive-Mode-Locking without saturable
absorber \cite{JMB-PRL-06}, though here it is operated in a completely
different regime.

\begin{figure*}[ht]
\centering{} \includegraphics[width=1\textwidth]{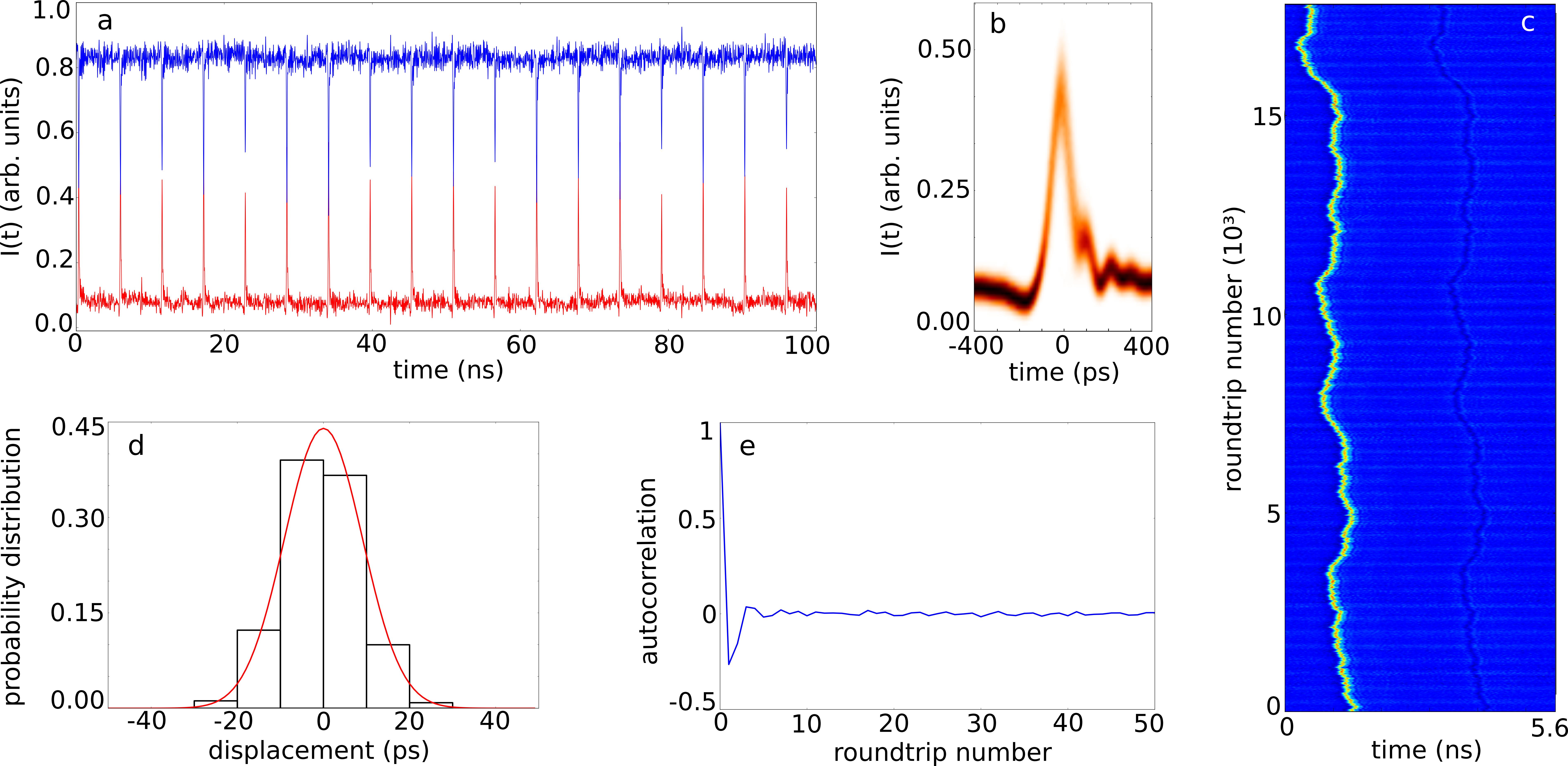}
\caption{Single Dissipative Soliton regime. Panel a): Time signal of the Y
(blue) and X (red) outputs when the VCSEL is submitted to XPR (rate
$1.4\,$\%) and PSF (rate $0.6\,$\%). $J=3.0\,$mA, $\tau_{f}=5.6\,$ns,
$\tau_{r}=8.4\,$ns. The double trace acquisition limits the sampling
rate at 100 GS/s Panel b): Histogram in color grade persistence resulting
from the superposition of $18\times10^{3}$ pulses in the X polarization
obtained using the same time trace as in b). Panel c): Space-time
representation diagram of a single DS obtained using the same time
trace as in b). The shadow to the right of the pulse is almost invisible
in the time trace and corresponds to a small inverted pulse (i.e.
upward in Y and downward in X) that appears at time $\Delta\tau=\tau_{r}-\tau_{f}=2.8\,$ns.
Panel d): Probability distribution of the residual $R_{n}=\Delta_{n}-\Delta_{n-1}$
calculated from the trace in a) with $\Delta_{n}$ the position of
the pulse at round-trip $n$ in c). Panel e) Autocorrelation of $R_{n}$.
\label{fig:mlocking}}
\end{figure*}
For properly chosen parameters (see Methods~\ref{sub:Setup-and-VCSEL}),
the polarization resolved outputs of the VCSEL exhibit a train of
pulses separated by $\tau_{f}=5.6\,$ns, as shown in Fig.~\ref{fig:mlocking}a.
In the X polarization component, the pulses are upward over a low
intensity background while they are downward from a high intensity
level in Y. The pulse duration is 80 ps (FWHM) and it is not fully
resolved due to the bandwidth limitation of the detector used (8 GHz).
These polarization pulses appear anti-correlated and the corresponding
total intensity time trace is almost constant, thus revealing the
vectorial character of this dynamics. We interpret these polarization
pulses as vectorial DS which travel back and forth in the external
cavity and get regenerated at each round-trip when interacting with
the VCSEL.

Dissipative solitons are required to maintain their shape throughout
the number of round-trips covered, though inevitable noise sources
(namely spontaneous emission in the laser, detector shot noise, mechanical
and thermal stability of the experimental set-up) can blur this ideal
picture, as shown in \cite{LCK-NAP-10}. We assessed the self-similarity
of the pulses by performing a statistical analysis of an $100\,\mu$s
long time series spanning over $18\times10^{3}$ pulses. The distribution
of the maxima of the pulses exhibits a Gaussian shape with a standard
deviation of $\sim10\,\%$ of the average peak intensity. By using
the peak of each pulse as a time reference, we superposed the $18\times10^{3}$
waveforms in Fig.~\ref{fig:mlocking}b. This reveals that, regardless
of the peak intensity fluctuations, the shape of the pulse remains
robust and stable, thus supporting our interpretation of the pulses
in terms of vectorial DS. 

The dynamics within an external cavity can be usefully described in
terms of space-time like diagrams where the time trace is folded over
itself at intervals $\tau_{f}$, so that the round-trip number becomes
the pseudo-time discrete variable while the pseudo-space variable
corresponds to the timing of the vectorial DS modulo $\tau_{f}$ \cite{AGL-PRA-92,GP-PRL-96}.
This representation pictures the soliton position $\Delta_{n}$ within
the external cavity as a function of the round-trip number \cite{JEM-NAP-13}.
When applied to the X polarization trace of Fig.~\ref{fig:mlocking}a
it reveals that $\Delta_{n}$ fluctuates noticeably over a typical
time scale of the order of a hundred of cycles, suggesting that the
noise present in the system acts upon the soliton as a Langevin force
over a free particle, leading to a Brownian-like motion of the soliton
within the external cavity as a function of the round-trips covered.
Indeed, the analysis of the time series for $\Delta_{n}$ indicates
that this variable is described by a first order autoregressive model,
$\Delta_{n}=\Delta_{n-1}+R_{n}$, with $R_{n}$ a random term which
is distributed as a Gaussian of zero mean and a standard deviation
$\sigma_{R}=11$ ps, as shown in Fig.~\ref{fig:mlocking}d. Moreover,
the autocorrelation of $R_{n}$ (Fig.~\ref{fig:mlocking}e) falls-off
very rapidly (a few round-trips) indicating that the sequence of $\Delta_{n}$
corresponds to the regular sampling of an one-dimensional stochastic
process $d_{t}\Delta=\xi\left(t\right)$ where $\xi(t)$ is a Gaussian,
slightly colored noise. The analogy between unidimensional Brownian
motion of a free particle and the drift of the DS position within
the cavity can be traced back to the temporal translational invariance
of our autonomous dynamical system \cite{BRL-PRE-03,KVT-PRA-08}.

\begin{figure}[ht]
\begin{centering}
\includegraphics[clip,width=1\columnwidth]{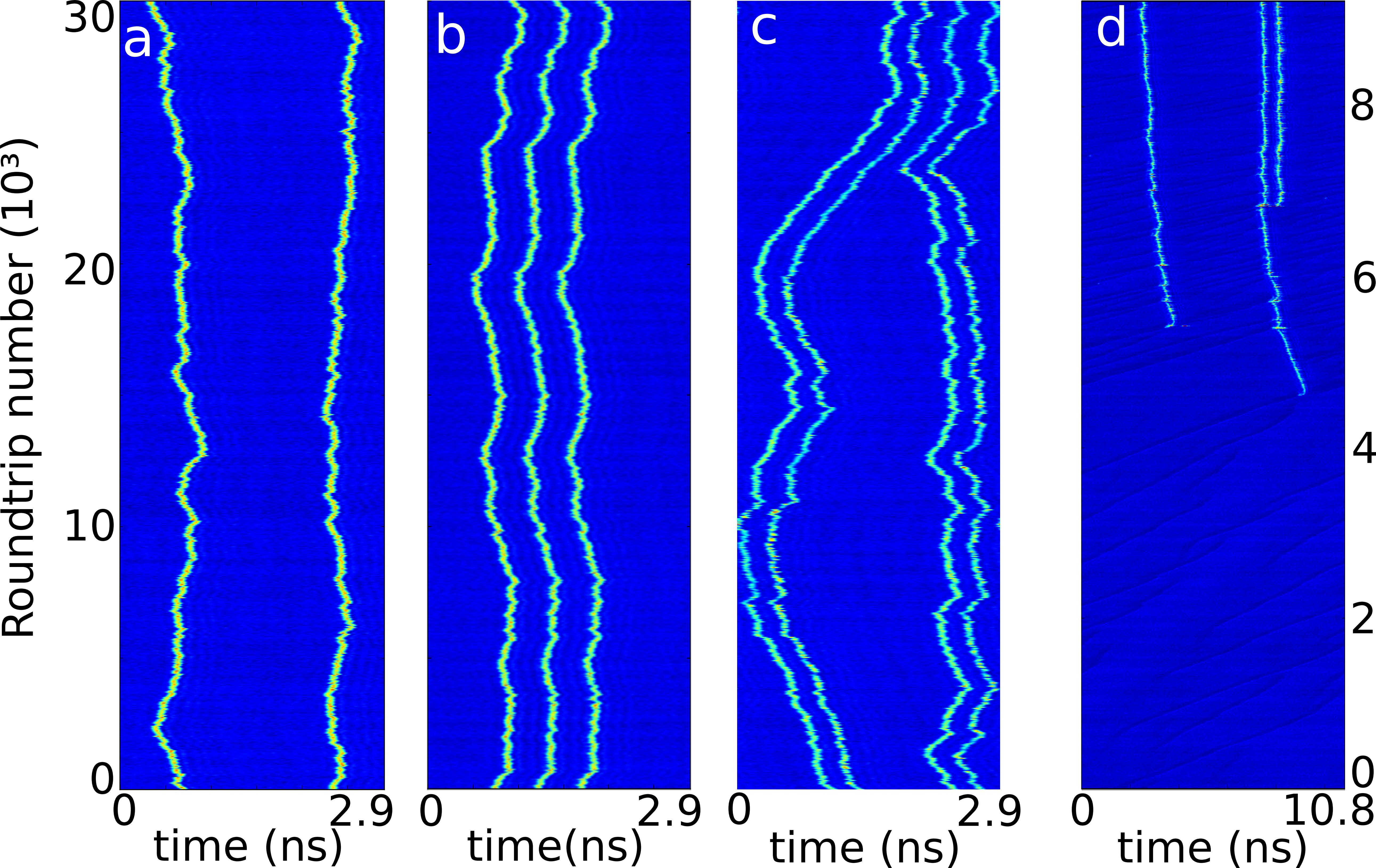}
\par\end{centering}

\caption{Space-time like diagrams of different states with multiple solitons
coexisting within the external-cavity. a) Two independent DS, b) a
3-DS molecule separated by a time $\Delta\tau$, c) two independent
2-DS molecules that form a 4-DS molecule at round-trip $12\times10^{3}$
separate and bind again at round-trip $23\times10^{3}$. In (a-c)
$\tau_{f}=2.9\,$ns and $\tau_{r}=3.3\,$ns. In d) we represent a
similar dynamics yet for longer delays $\tau_{f}=10.8\,$ns and $\tau_{r}=11.13\,$ns;
here several vectorial DS spontaneously nucleate due to noise induced
fluctuations at round-trip $4.5\times10^{3}$, $5.2\times10^{3}$
and $7\times10^{3}$, thus evidencing the multi-stability with the
CW solution. While the first and second DS are independent, the third
one forms a molecule with the second at a bonding distance of $600\,$ps,
i.e. at a distance\emph{ twice} the nominal separation, i.e. $2\Delta\tau$.
All the other parameters are as in Fig.~\ref{fig:mlocking}.\label{fig:indep}}
\end{figure}

If $\tau_{f}$ is large enough compared with the size of a DS, several
localized structures can be hosted within the external cavity and
their interaction may be studied. Then, the noise induced motion of
the DS becomes a tool for discriminating between independent DS and
DS bound states or molecules. We show in Fig.~\ref{fig:indep}a-c
the space-time diagrams for different ensembles of coexisting DS.
We remark that these different realizations have been obtained for
the same parameter values; the system may evolve from one situation
to another in response to perturbations or parameter sweeps, in the
latter case displaying a high degree of hysteresis. Panel a) shows
two DS whose noise-induced trajectories are uncorrelated, thus evidencing
their independence. Panel b) shows a molecule of three DS; while the
evolution of the ensemble is stochastic, the separation between the
DS remains constant at $480\,$ps, which corresponds to $\Delta\tau=\tau_{r}-\tau_{f}$
(Supplementary Section~\ref{sub:Complexes-and-bond}, Fig.~\ref{histo}a,b).
We depict in Fig.~\ref{fig:indep}c the binding and unbinding between
two molecules, each one formed by two DS. At round-trip $12\times10^{3}$,
the two molecules approach and form a bound state with four peaks
that afterwards moves as a single structure and subsequently unbind
and bind again at latter times. The situations reported in Fig.~\ref{fig:indep}a-c
are a small sample of the multiple situations in terms of DS number
and organization (molecules of variable number of DS) that coexist
for the same values of the parameters. For $\tau_{f}=2.9\,$ ns, molecules
composed by a larger number of DS can be observed (Supplementary Section~\ref{sub:Complexes-and-bond},
Fig.~7a), the largest one being an 8-DS molecule filling the entire
round-trip and forming a DS \textquotedbl{}crystal\textquotedbl{}\textsl{.}\textsl{\emph{
Increasing the size of the external cavity allows placing a higher
number of independent DS and larger molecules (Supplementary }}Section~\ref{sub:Complexes-and-bond},
Fig.~7\textsl{\emph{b,c)}}\textsl{. }In Fig.~\ref{fig:indep}d we
show how different vectorial DS may nucleate spontaneously from the
CW solution, i.e. from a situation with zero solitons. Fig.~\ref{fig:indep}d
suggests that DS can be addressed inside the cavity by a proper perturbation
pulse and used as bits for information processing, as reported in
\cite{LCK-NAP-10}.

\section{Theory\label{theory}}

\begin{figure}[ht]
\begin{centering}
\includegraphics[bb=0bp 0bp 412bp 152bp,clip,width=0.45\textwidth]{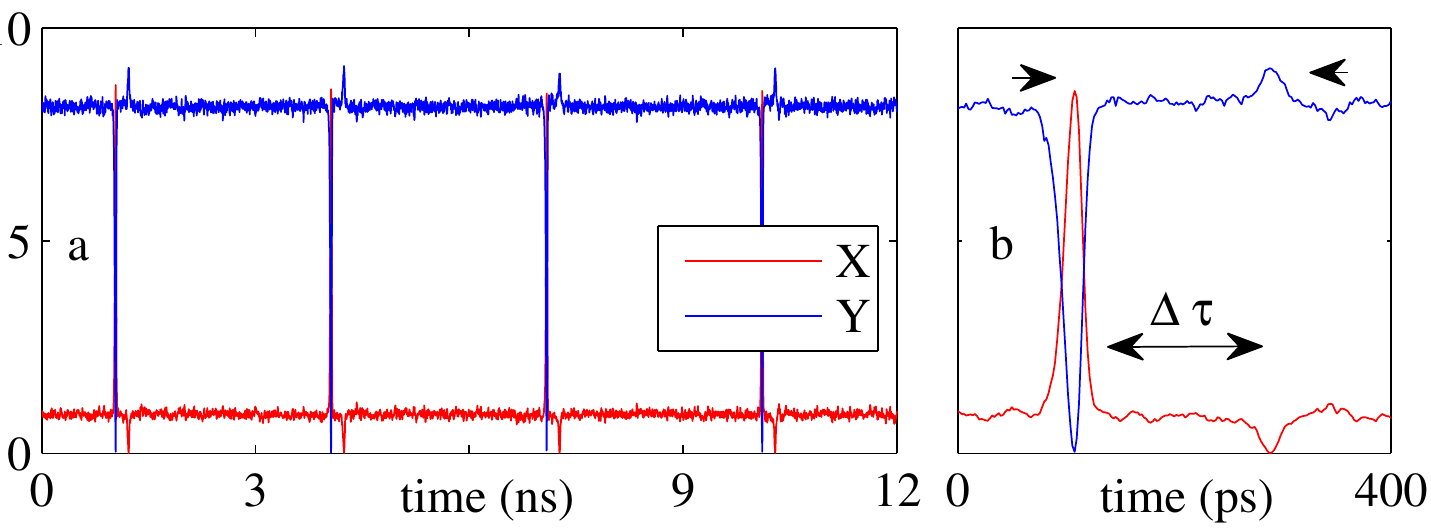} 
\par\end{centering}

\begin{centering}
\includegraphics[bb=0bp 0bp 524bp 338bp,clip,width=0.48\textwidth]{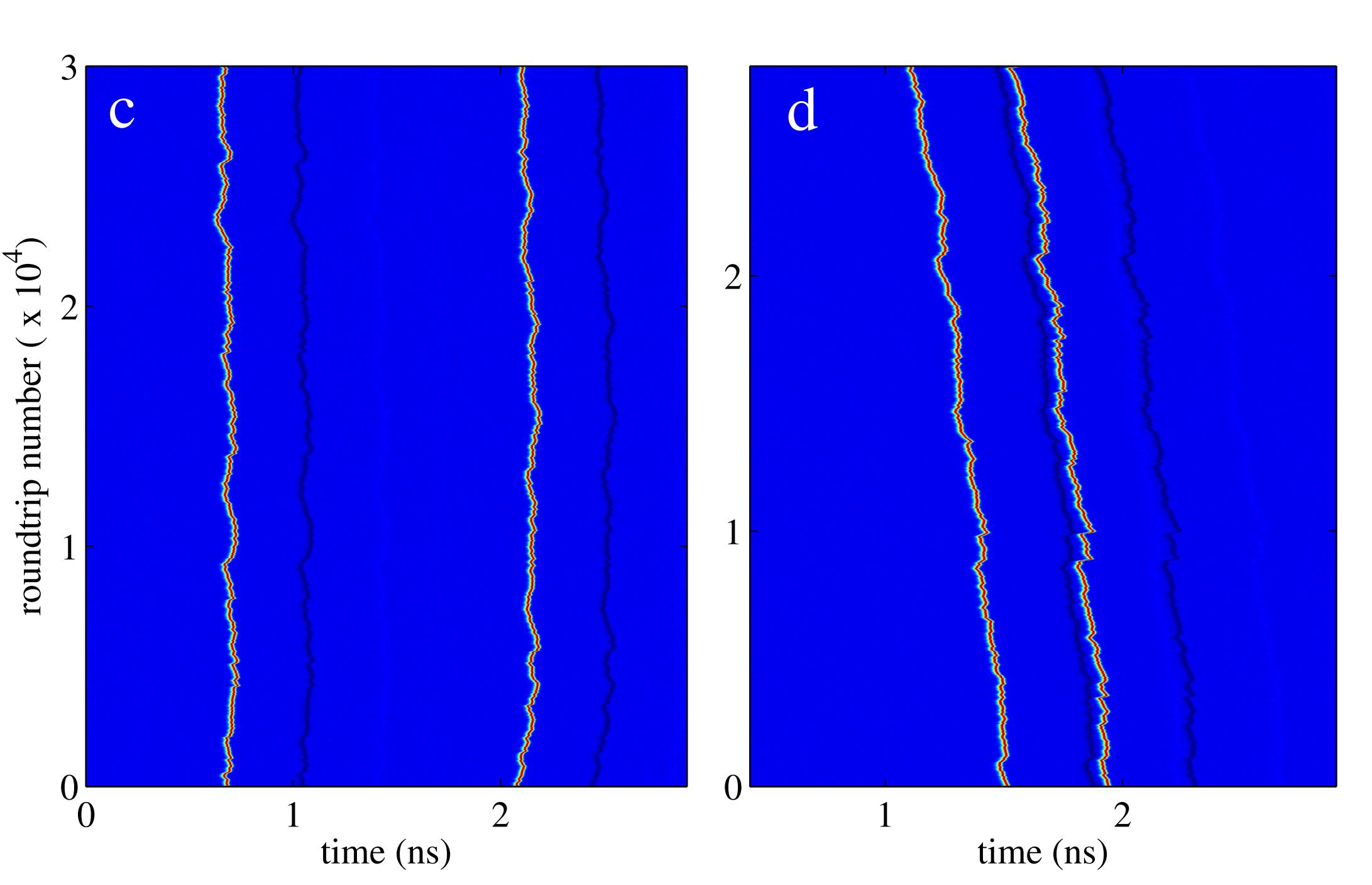}
\par\end{centering}

\caption{Theoretical temporal trace for a single vectorial DS (a) created by
perturbing the phase of PSF (see Methods~\ref{sub:Theoretical-model})
and (b) zoom on the pulse detail: the anti-phase dip is followed by
a small inverted kink after a time delay $\Delta\tau$ corresponding
to the re-injection of Y into X after a longer time $\tau_{r}$. Other
exponentially decreasing replica of this secondary kink (not visible)
follow at time intervals $n\Delta\tau$. (c-d) Folded space-time representation:
the shadow following the DS corresponds to the inverted dip at $\Delta\tau$.
In (c) the second DS was created far from the first one yielding two
independent objects as evidenced by their uncorrelated motion. Conversely,
if the second DS is nucleated at some precise closer distance, one
obtains a bound state (d) where the two solitons exhibit correlated
motions. The period of the solutions is slightly superior to $\tau_{f}$,
such secular drift being due to the finite response time of the VCSEL.
\label{theo1_tracemol} }
\end{figure}

In order to explain the experimental results, we use the Spin-Flip
Model \cite{SFM-PRA-95}, suitably modified for incorporating the
effects of gain saturation, PSF and XPR. The choice of parameter values
was guided by the experimental situation (see Methods~\ref{sub:Theoretical-model}).
Fig.~\ref{theo1_tracemol}a shows a dynamical state for the 1-DS
case in good agreement with the experiment: the intensity of each
linearly polarized component displays localized anti-phase pulsations
separated by the PSF time delay, followed by a small inverted kink
after a time $\Delta\tau$, see Fig.~\ref{theo1_tracemol}b. The
presence of this inverted kink in the experimental data can be appreciated
in Fig.~\ref{fig:mlocking}c as a dark shadow following the DS at
a distance $\Delta\tau$. We predict an almost $100\,$\% anti-phase
as well as pulse-widths of the order of $\sim35\,$ps which is compatible
with the experimental results taking into account the bandwidth limitations.
Importantly, such regime coexists with a CW solution and in-between
pulses (or in the absence of them), the emission consists in a quasi-linearly
polarized mode whose orientation is neither X nor Y. The polarization
orientation is governed by a complex interplay between the dichroism,
the birefringence, the PSF and the XPR rates as well as the two delays,
see \cite{JMG-PRA-14} for details. Typically, the suppression ratio
between $I_{y}/I_{x}$ can be tuned between $20$ and $5$. 

In addition to isolated DS, we reproduce the coexistence of multiple
independent solitons, see Fig.~\ref{theo1_tracemol}c. Additional
DS can be written at arbitrary positions without perturbing the already
present localized structures. The bound states are accounted for as
well, which can be appreciated in Fig.~\ref{theo1_tracemol}d. We
explain the existence of a specific binding distance to the small
kink generated in the wake of the vectorial DS by the replica of the
main DS impinging the VCSEL a second time after a time interval $\Delta\tau$,
in agreement to experimental results of Fig.~\ref{fig:mlocking}c.
Notice that several exponentially decreasing replica at time $2\Delta\tau$,
$3\Delta\tau$ exist and generate a weak binding force as well as
other equilibrium distances for the molecules (Supplementary Section
\ref{sub:Fast-DS-and_mol}). Finally, for slightly different parameters
values, a more complicated situation exists in which the background
CW solution weakly oscillates in anti-phase for X and Y at a frequency
close to the birefringence. This induces additional binding distances
whose separation is $\sim120\,$ps around the main resonances $n\Delta\tau$
. Such complexity is typical of the interaction between vectorial
DS, see e.g. \cite{SFA-PRL-00}, and will be reported elsewhere.

\begin{figure}[ht]
\begin{centering}
\includegraphics[bb=0bp 0bp 483bp 344bp,clip,width=0.49\textwidth,height=0.38\textwidth]{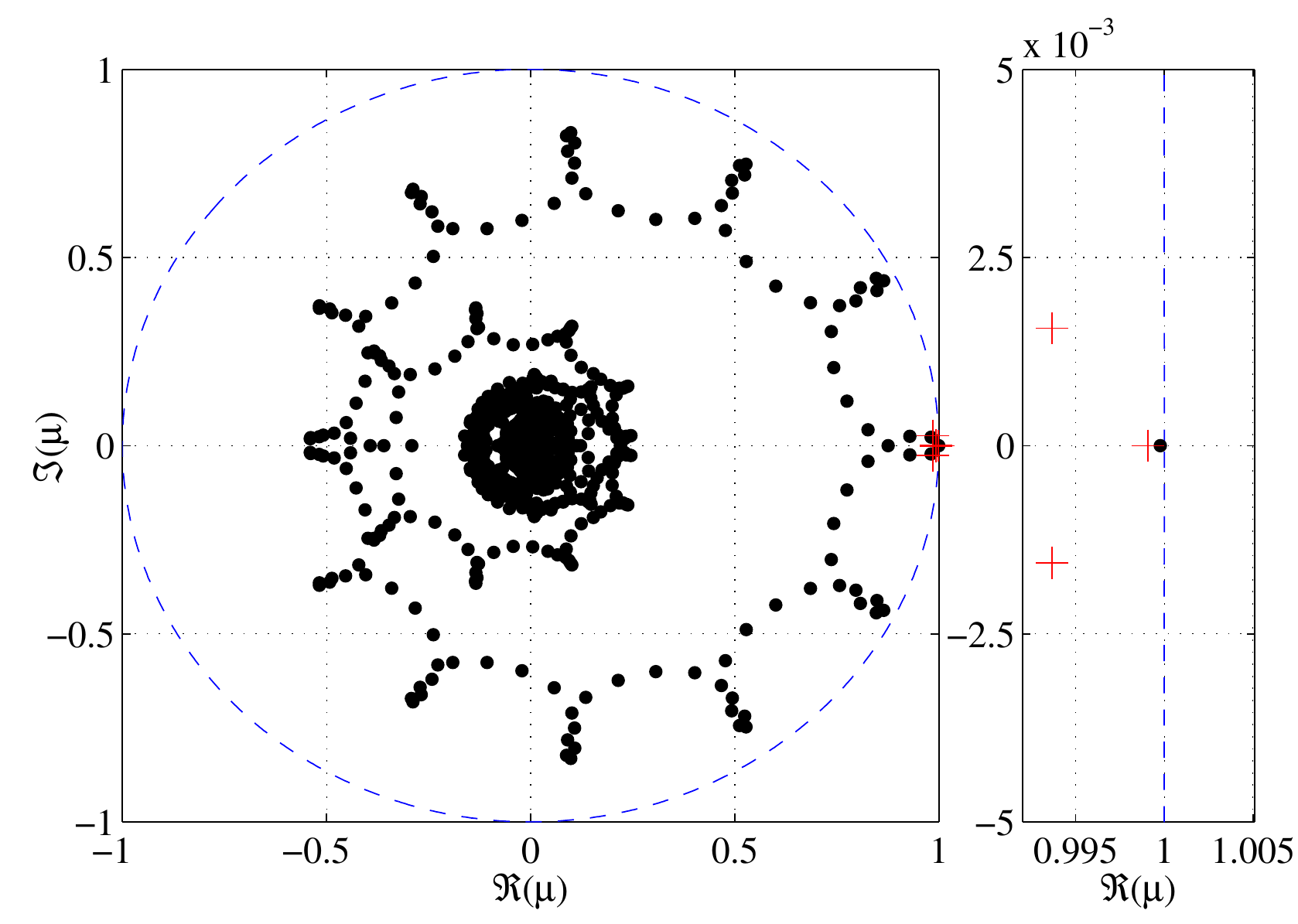}
\par\end{centering}

\caption{Floquet multipliers $\mu$ (black circles) of the single DS solution
showed in Fig.~\ref{theo1_tracemol}a. All the multipliers have a
modulus smaller than unity, as expected for a stable solution. A zoom
in the vicinity of $\mu=1$ allows finding a single multiplier as
anticipated for a periodic solution in a time independent dynamical
system. For the case of $N=3$ DS, we represent for clarity only the
rightmost multipliers (red crosses): we notice the existence of not
one but \emph{three} quasi-degenerate Floquet multipliers having no
equivalent in the case of the single DS. This demonstrates the existence
of $N$ degenerate neutral mode that correspond to independent translational
motion of the various DS. The fact that the three multipliers are
not exactly equal to unity is a consequence of the ultra-weak residual
interactions between distant DS. Such residual interaction implies
that --- strictly speaking --- multiple DS are not totally independent
for any finite value of the delay. \label{theo2_floquet} }
\end{figure}

As a proof of the independence of the DS, we performed the analysis
of the Floquet multipliers considering the DS as periodic solutions
of a high dimensional dynamical system (Supplementary Section~\ref{sub:Floquet-analysis}).
The results of our analysis are depicted in Fig.~\ref{theo2_floquet}
for the case of one and three DS. With three DS, we notice in the
vicinity of $\mu=1$ the existence of not one but three quasi-degenerate
Floquet multipliers. We analyzed the eigenvectors associated with
these neutral modes and found that they correspond to relative translations
of each DS, further confirming, beyond the observation of correlated
or uncorrelated motion in Fig.~\ref{theo1_tracemol}c,d, their independence
and defining mathematically the concept of \emph{temporal localization}
within the framework of delayed systems. Interestingly, the analysis
of bound states yielded a single multiplier $\mu=1$ confirming that
the ensemble moves as a single entity. Finally, the presence of many
weakly damped oscillatory modes in Fig.~\ref{theo2_floquet} explains
the variations of height and the sensitivity of the system to noise.

\begin{figure*}[th]
\begin{centering}
\includegraphics[bb=0bp 0bp 409bp 350bp,clip,width=0.35\textwidth]{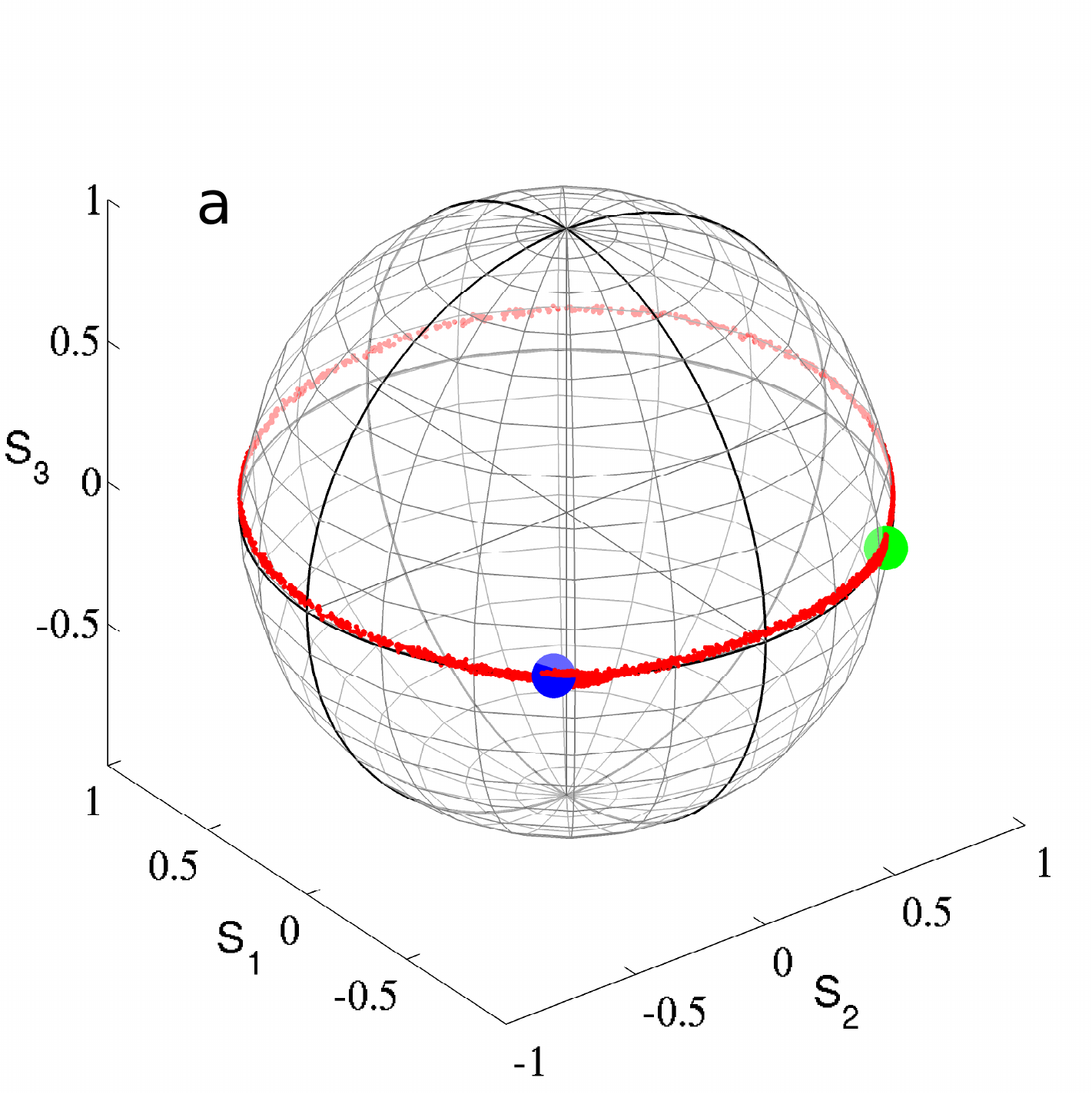}\includegraphics[bb=0bp 0bp 409bp 410bp,clip,width=0.3\textwidth]{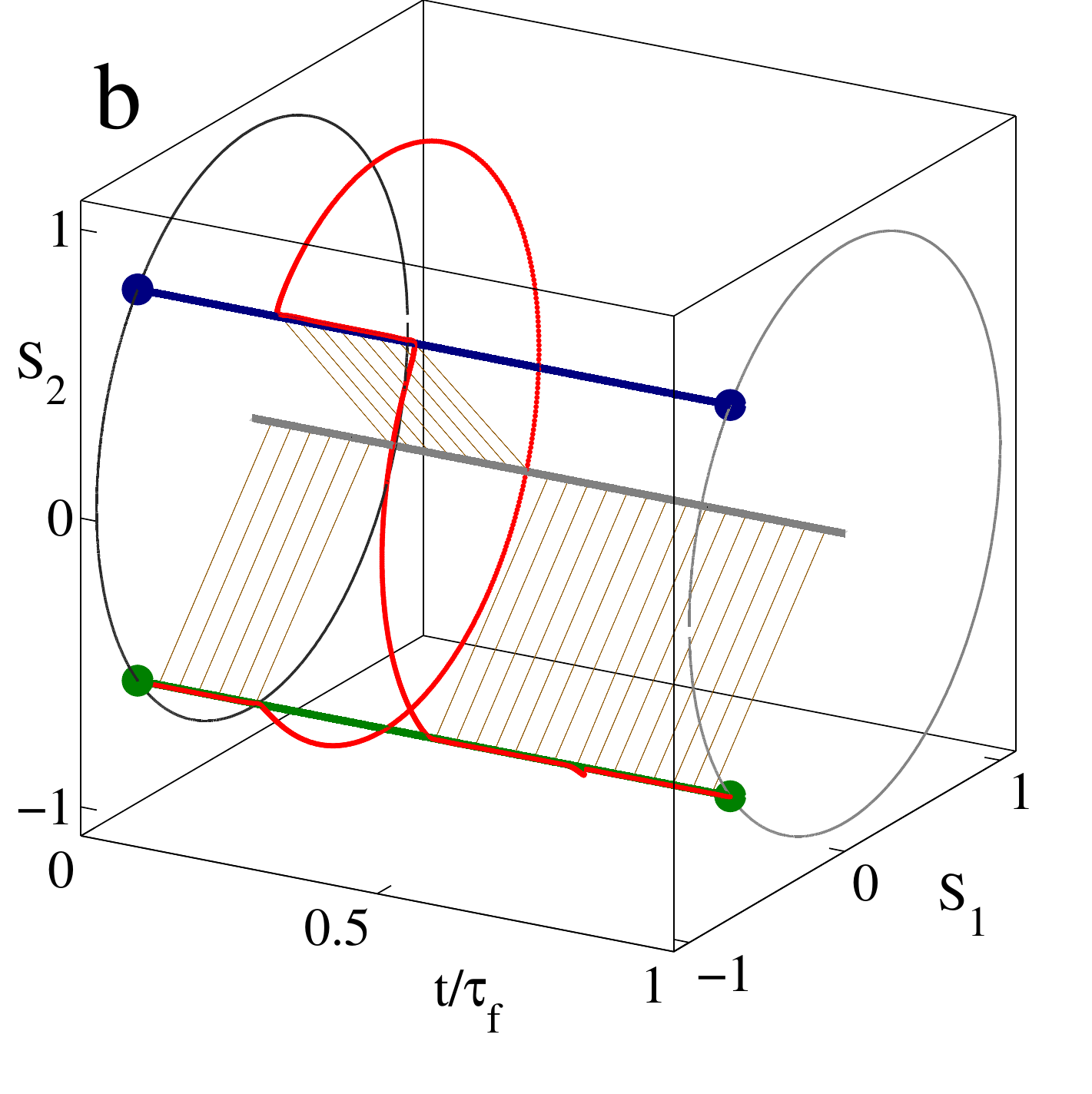}\includegraphics[bb=0bp 0bp 409bp 410bp,clip,width=0.3\textwidth]{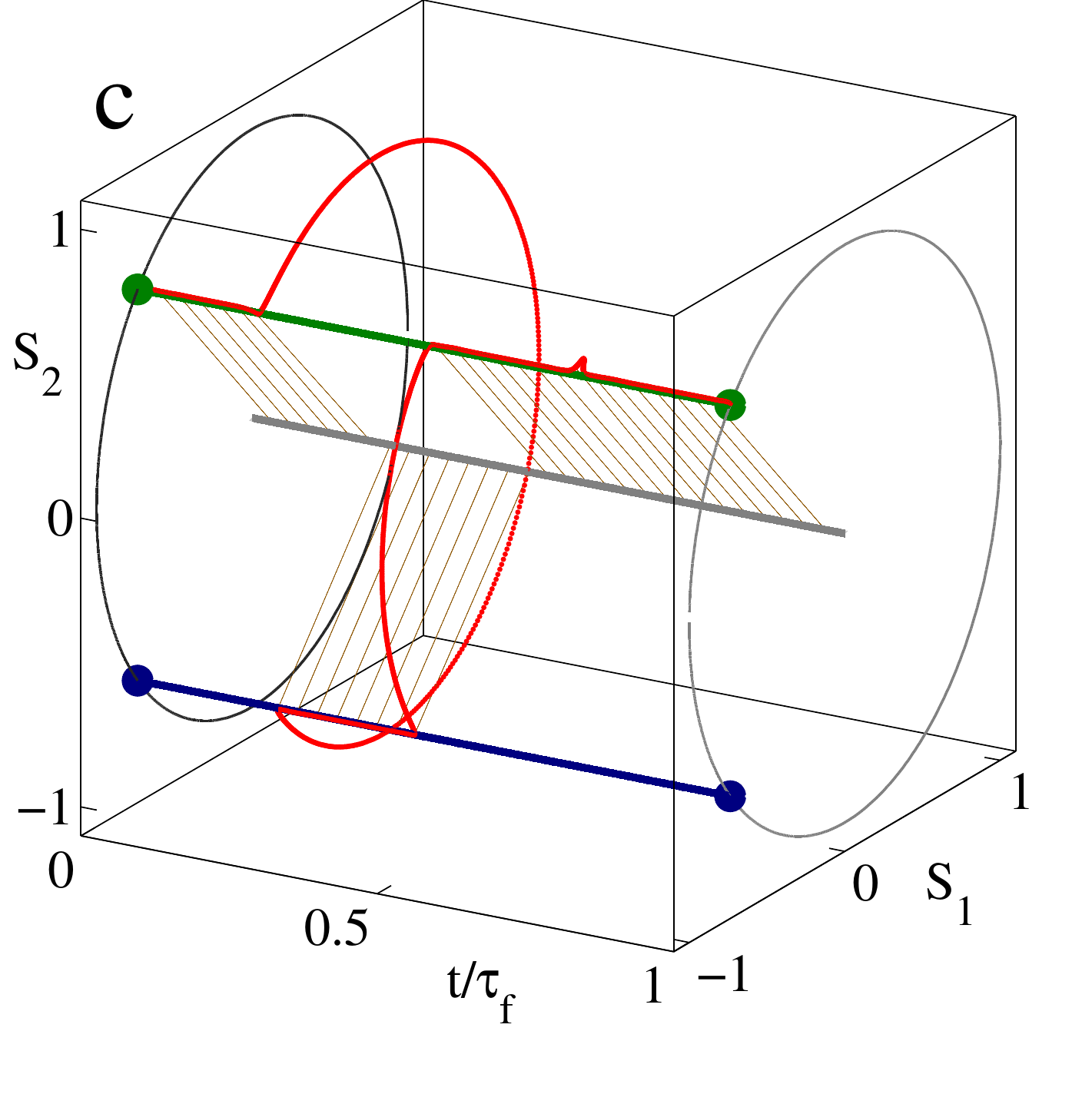} 
\par\end{centering}

\caption{Temporal trace using the normalized Stokes representation of the dynamics.
(a) Superposition of $10^{3}$ consecutive round-trips for a single
soliton. One notices that the polarization angle performs a full cycle,
but remains essentially close to the equatorial plane as indicated
by the weakness of the $S_{3}$ component. The position of the stable
fix point is denoted in green while the blue circle represents the
symmetrical polarization with respect to the Y axis of symmetry .
We recall that pure X and Y emission correspond to the equatorial
point $S_{1}=1$ and $S_{1}=-1$, respectively, and that $\pm45^{\circ}$
emission correspond to $S_{2}=\pm1$. A single orbit unfolded in time
is presented in (b). Here, one notices the existence of two plateaus
corresponding to the anti-phase dip followed by the small inverted
kink replica after a time $\Delta\tau$. A stable anti-kink was generated
in (c) by inverting the phase of the XPR, i.e. setting $\beta\rightarrow-\beta$.
\label{theo3_Stokes} }
\end{figure*}

A description based on the Stokes parameters for polarized light allows
interpreting the anti-phase dynamics as rotations along the equator
of the Stoke's sphere, thus unveiling the vectorial character of the
DS. We represent in Fig.~\ref{theo3_Stokes}a the temporal trace
corresponding to one DS. It is found that the orbit proceeds essentially
along the equator of the sphere $\left(\left|S_{3}\right|\lesssim0.1\right)$.
The system starts from the stable quasi-linearly polarized state represented
by a green circle and performs a full clockwise rotation to reach
the blue circle. Because these two polarizations are degenerate in
a representation based solely upon the intensity dynamics of X and
Y, one would think having reached the initial point. Yet, it is only
after receiving the second delayed perturbation after an additional
time delay $\Delta\tau$ that the polarization cycle is closed. Incidentally,
this explains why the secondary kink in Fig.~\ref{theo1_tracemol}a,b
is upward since one must pass through the pure Y emission state ($S_{1}=-1$)
during this secondary plateau. Such dynamics along the equator is
depicted in Fig.~\ref{theo3_Stokes}b where one can clearly identify
two plateaus corresponding to the aforementioned symmetrical polarizations. 

Such phase kinks are reminiscent of the Sine-Gordon phase equation
which is known to give topological kinks and anti-kinks solutions.
In our case, the phase must be interpreted as the orientation of the
polarization. By applying a multiple time scale to our model (Supplementary
Section~\ref{sub:Vectorial-phase-model} as well as \cite{JMG-PRA-14}
for details) one reduces the dynamics to the sole orientation angle
over the equatorial plane $\Phi$, such that $\left(S_{1},S_{2}\right)=\left(\cos\Phi,\sin\Phi\right)$.
The equation governing the dynamics of $\Phi$ reads 
\begin{eqnarray*}
\frac{\dot{\Phi}}{2} & = & \left(\alpha\gamma_{p}+\gamma_{a}\right)\sin\Phi+\bar{\eta}\sin\frac{\Phi^{\tau_{f}}}{2}\cos\frac{\Phi}{2}-\bar{\beta}\sin\frac{\Phi}{2}\sin\frac{\Phi^{\tau_{r}}}{2},
\end{eqnarray*}
with $\Phi^{\tau_{f,r}}=\Phi\left(t-\tau_{f,r}\right)$ the delayed
arguments. This reduced model clarifies which parameters control the
dynamics: the magnitude of the effective PSF and XPR rates dressed
by the $\alpha$ factor, $\left(\bar{\eta},\bar{\beta}\right)=\left(\eta,\beta\right)\sqrt{1+\alpha^{2}}$,
the dichroism $\gamma_{a}$ and the birefringence $\gamma_{p}$. The
results of our reduced model is in quantitative agreement with the
results of the full model and was indeed used to generate Fig.~\ref{theo3_Stokes}b.
Elaborating upon these predictions, we reduced $\Delta\tau$ in order
to obtain a single uninterrupted cycle and we were able to find DS
as short at $25\:$ps, both in the full and the simplified model (Supplementary
material Section~\ref{sub:Fast-DS-and_mol}). Also, exploiting the
symmetry property $\left(\Phi,\beta\right)\rightarrow\left(-\Phi,-\beta\right)$,
we deduce the existence of topological anti-kink solutions, which
are depicted in Fig.~\ref{theo3_Stokes}c. Notice that the kink and
the anti-kink can not be separated in our measurements. More complicated
kinks and anti-kinks that do not correspond to entire rotations also
exist and will be the topic of further studies. 

In summary, we demonstrated the existence of vectorial dissipative
solitons and molecules in a VCSEL enclosed in a double external cavity.
A simple theoretical model was found to reproduce the main experimental
features and the pulse shadow was identified as the pinning mechanism
allowing the creation of molecules. The proof of independence was
given experimentally by the study of the DS random motion and theoretically
by the analysis of the Floquet multipliers. We interpreted the DS
as rotations along the equatorial plane of the Stokes sphere and a
multiple time-scale analysis allowed us to reduce the dynamics to
a single delayed equation for the polarization orientation.

\section*{Acknowledgements }

J.J. acknowledges financial support from the Ramón y Cajal fellowship
and useful discussions with B. Krauskopf. J.J. and S.B. acknowledge
financial support from project RANGER (TEC2012-38864-C03-01) and from
the Direcció General de Recerca, Desenvolupament Tecnològic i Innovació
de la Conselleria d'Innovació, Interior i Justícia del Govern de les
Illes Balears co-funded by the European Union FEDER funds. M.G. acknowledge
discussions with G. Giacomelli and C. Miniatura. The INLN group acknowledges
funding of Région \textquotedbl{}Provence-Alpes-Cote d'Azur\textquotedbl{}
with the \textquotedbl{}Projet Volet Générale 2011 : Génération et
Détection des Impulsions Ultra Rapides (GEDEPULSE)\textquotedbl{}
and project ANR \textquotedbl{}OPTIROC\textquotedbl{} and project
ANR - Jeunes Chercheurs \textquotedbl{} MOLOSSE \textquotedbl{}.

\subsection*{Author Contributions }

M. Marconi performed the experimental characterization under the supervision
of M. Giudici and assisted by S. Barland. M. Marconi, S. Balle and
M. Giudici performed the statistical analysis of the experimental
data. J. Javaloyes developed the theoretical model, the numerical
analysis and wrote the manuscript together with S. Balle and M. Giudici.
All the authors participated to the interpretation of the results.

\subsection*{Competing Interests }

The authors declare that they have no competing financial interests.

\subsection*{Materials \& Correspondence }

Correspondence and requests for materials should be addressed to Massimo
Giudici (email: massimo.giudici@inln.cnrs.fr).

\section{Methods}

\subsection{Setup and VCSEL details\label{sub:Setup-and-VCSEL}}

Two different lasers from ULM-photonics lasing at $850\,$ nm (ULM.850-PMTNS46FOP)
have been used. They are single transversal mode lasers with a suppression
ratio larger than $10\,$dB at the rated power of $1\,$mW. Both emit
linearly polarized light and exhibit a birefringence around $7.5\,$GHz.
Their low dichroism and birefringence helps to maximize the effects
of PSF and XPR. Their threshold is around $0.5\,$mA with the bluest
polarization mode, that we call Y, appearing at threshold. Both devices
exhibit a small dichroism and polarization switching can be implemented
by injecting an external linearly polarized beam of low power ($<\text{100\,}$nW).
It is important to underline that the linear polarization component
selected by the PSF cavity and used for XPR must be the dominant component
appearing at threshold of the solitary laser which is Y, the bluest
one, in our case. The DS are obtained by choosing some specific values
of the system parameters. Typical feedback rate for this regime is
around $0.5\,\%$, while the XPR rate is about $1-2\,\%$. Pumping
currents are $2.6\,\mathrm{mA}<J<3.1\,\mathrm{mA}$. In this experimental
analysis we explored $\tau_{f}$ values ranging from 1.3~ns up to
10~ns. The first value is the lowest limit possible with our setup,
while the second one is just the maximum value explored. The detection
set-up consists of two $8\,$GHz detectors combined with a $33\,$GHz
scope (Tektronix DPO73304D).

\subsection{Theoretical model\label{sub:Theoretical-model}}

In order to theoretically analyze the behavior of the VCSEL and compare
with the experimental results, we use the Spin-Flip Model (SFM) \cite{SFM-PRA-95},
suitably modified for incorporating the effects of both polarization
selective feedback and crossed-polarization re-injection. We adopt
a mixed description in terms of linearly polarized components of the
field, X and Y, where PSF and crossed-polarization re-injection are
easily expressed, and circularly polarized components of the field,
$E_{\pm}=(X\pm iY)/\sqrt{2}$, where the SFM is naturally expressed.
In this framework, the dynamics of the system is written as 
\begin{eqnarray}
\dot{X} & = & \left(1+i\alpha\right)\left[\frac{G_{+}E_{+}+G_{-}E_{-}}{\sqrt{2}}-X\right]-zX\nonumber \\
 & + & \beta e^{-ia}Y(t-\tau_{r})\label{eq:X}\\
\dot{Y} & = & \left(1+i\alpha\right)\left[\frac{G_{+}E_{+}-G_{-}E_{-}}{i\sqrt{2}}-Y\right]+zY\nonumber \\
 & + & \eta e^{-i\Omega}Y(t-\tau_{f})\label{eq:Y}\\
T\dot{D}_{\pm} & = & \mu-D_{\pm}-G_{\pm}\vert E_{\pm}\vert^{2}\mp\frac{\gamma_{J}}{\gamma_{e}}\left(D_{+}-D_{-}\right),\label{eq:D}
\end{eqnarray}
 where $\alpha$ is Henry's linewidth enhancement factor, $D_{\pm}$
are the scaled carrier densities in each spin channel and $G_{\pm}=D_{\pm}\left(1-\varepsilon\vert E_{\pm}\vert^{2}\right)$
is the gain for each circularly polarized field component including
gain saturation. The terms $Y(t-\tau_{f})$ and $Y(t-\tau_{r})$ in
the evolution equations for the X and Y-components describe the effects
of PSF and of XPR, which have strengths $\eta$ and $\beta$, phases
$\Omega$ and $a$, and time delays $\tau_{f}$ and $\tau_{r}$, respectively.

In equations (\ref{eq:X}-\ref{eq:D}), time has been scaled to the
cavity decay rate $\kappa$, $T=\kappa/\gamma_{e}$ represents the
scaled carrier lifetime and $\gamma_{J}$ is the decay rate of the
spin-difference. The scaled density of carriers injected per unit
time into the active region due to the bias current normalized to
threshold is $\mu$. In addition, we defined $z=\left(\gamma_{a}+i\gamma_{p}\right)/\kappa$
where $\gamma_{a}$ (resp. $\gamma_{p}$) describes the linear dichroism
(resp. birefringence) of the cavity. Finally, we have added to the
time evolution in Eqs.~(\ref{eq:X}-\ref{eq:D}) independent Langevin
sources describing noise due to spontaneous emission and current fluctuations
\cite{MMS-PRA-01} with variance $\xi\sim10^{-3}$. We assume typical
values $\alpha=2$, $\varepsilon_{g}=2\times10^{-2}$,$\gamma_{a}=0$,
$\gamma_{p}=5\times10^{-2}$, $\mu=10$, $T=500$ and $\gamma_{J}=60\gamma_{e}$
which --- taking $\kappa^{-1}=2\,$ps --- correspond to a carrier
lifetime $\gamma_{e}^{-1}=1\,$ns, a spin-difference decay time of
$\gamma_{J}^{-1}=16.6\,$ps, a frequency splitting $\gamma_{p}\kappa/\pi\sim8\,$GHz
and a relaxation oscillation frequency $\nu_{r}=\left(2\pi\right)^{-1}\sqrt{2\left(\mu-1\right)\gamma_{e}\kappa}\sim15\,$GHz.
The values of the PSF and XPR parameters used in all theoretical figures
are $\eta=0.09$ and $\beta=0.06$ while $\Omega=a=0$. The time delays
in Fig.~\ref{theo1_tracemol}a,b, Fig.~\ref{theo2_floquet} and
Fig.~\ref{theo3_Stokes} are $\tau_{f}=1500$ and $\tau_{r}=1600$
and correspond to $3\,$ns and $3.2\,$ns. We took $\tau_{r}=1700$
(i.e. $3.4\,$ns) in Fig.~\ref{theo1_tracemol}c,d in order to better
visualize the shadow of the pulse and $\tau_{r}=1900$ (i.e. $3.6\,$ns)
in Fig.~\ref{theo3_Stokes}b,c in order to clarify the existence
of the plateaus. For completeness, we recall that the normalized Stokes
parameters read 
\begin{eqnarray*}
S_{0}=\left|X\right|^{2}+\left|Y\right|^{2} & \;,\  & S_{1}=\left(\left|X\right|^{2}-\left|Y\right|^{2}\right)/S_{0}\\
S_{2}=2\Re\left(XY^{\star}\right)/S_{0} & \;,\; & S_{3}=-2\Im\left(XY^{\star}\right)/S_{0}.
\end{eqnarray*}

\subsection{Numerical simulations\label{sub:Numerical-simulations}}

Eqs.~(\ref{eq:X},\ref{eq:D}) were numerically integrated with a
fourth-order Runge-Kutta method with constant step size ($h=10^{-2}$)
\cite{NR-BOOK}. The delayed contributions in Eqs.~(\ref{eq:X},\ref{eq:Y})
demand a special care. To advance the solution with a step $h$ from
$t_{n}=nh$ to $t_{n+1}$, the Runge-Kutta algorithm requires the
values of $Y\left(t-\tau_{f,r}\right)$ at intermediate points $t_{n+1/2}$.
These are not known and must be interpolated from past values with
an order of approximation consistent with that of the algorithm of
integration. Therefore, besides keeping memory of the past values
of $Y$ we also retain the past values of the time derivative $\dot{Y}\left(t\right)$.
Such a method allows building a third order Hermite polynomial approximation
for $Y(t)$ between the time $\left(t_{n}-\tau_{f,r}\right)$ and
$\left(t_{n+1}-\tau_{f,r}\right)$. By evaluating this interpolating
polynomial at $\left(t_{n+1/2}-\tau_{f,r}\right)$, we ensure an overall
fourth order accuracy. Finally, the stochastic noise contribution
in Eqs.~(\ref{eq:X}-\ref{eq:D}) is added after the deterministic
step by simply using the Euler method \cite{NR-BOOK} and scaling
$\xi$ by $\sqrt{h}$.

Solitons can be generated numerically by starting from a random initial
condition around the off solution; during the transient regime, several
spikes corresponding to the so-called relaxation oscillations are
generated which can result in one or several DS in the asymptotic
regime. A more controlled approach is however more appropriate to
generate molecules. We use as an initial condition the CW mode, i.e.
the solution with 0-DS . From this state, by inverting the feedback
phase $\Omega\rightarrow\Omega+\pi$ one can generate polarization
slips that will eventually stabilize as vectorial DS after a few tens
round-trips in the cavity. Notice that setting the phase back to its
original value, even after a short time interval generates another
Soliton and if the interval is too short no DS is generated. From
this newly found periodic solution, additional perturbations can be
applied in order to generate multiple independent solitons or soliton
molecules if the perturbation is properly timed.

\bibliographystyle{unsrt}
\bibliography{full}

\section{Supplementary Material}

\subsection{Complexes and bond distances\label{sub:Complexes-and-bond}}

\begin{figure}[ht]
\centering{}\includegraphics[bb=0bp 0bp 1214bp 781bp,clip,width=0.9\columnwidth]{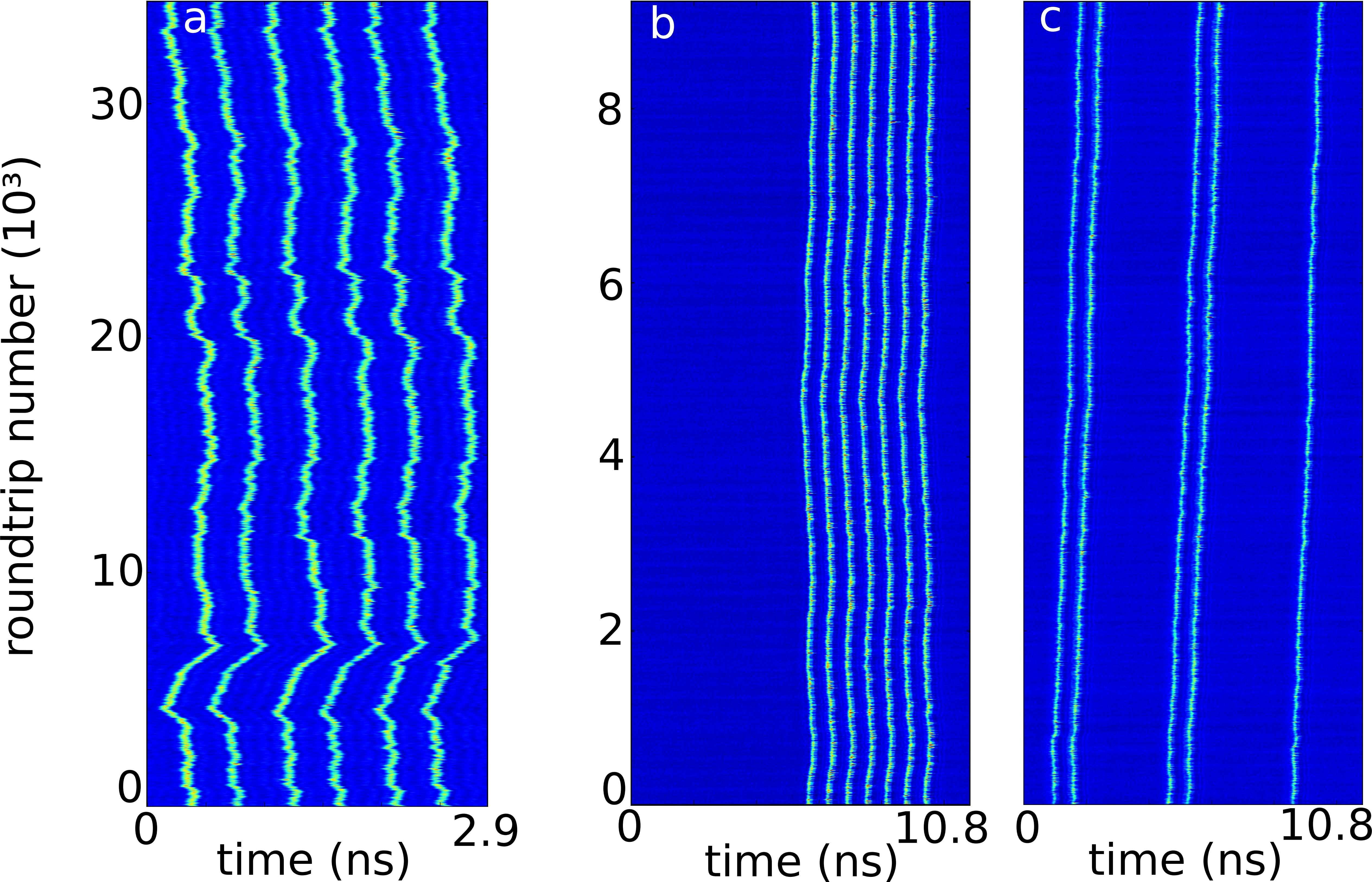}\caption{Panel a): Space-time like diagram of a 6-DS molecule obtained for
the same parameters of Fig.~\ref{fig:indep}a ($\tau_{f}=2.9\,$ns
and $\tau_{r}=3.3\,$ns). By increasing $\tau_{f}$, a larger number
of independent DS and molecules with a higher number of elements can
fit into the cavity. Panel b): Space-time like diagram of a molecule
of 7-DS evolving in a limited portion of an external cavity having
$\tau_{f}=10.8\,$ns and $\tau_{r}=11.1\,$ns. Panel c): Another situation
obtained for the same parameter as panel b) showing three independent
DS: a single DS and two 2-DS molecules. \label{molecules} }
\end{figure}

\begin{figure*}[th]
\centering{}\includegraphics[bb=0bp 0bp 1712bp 1036bp,clip,width=0.49\textwidth]{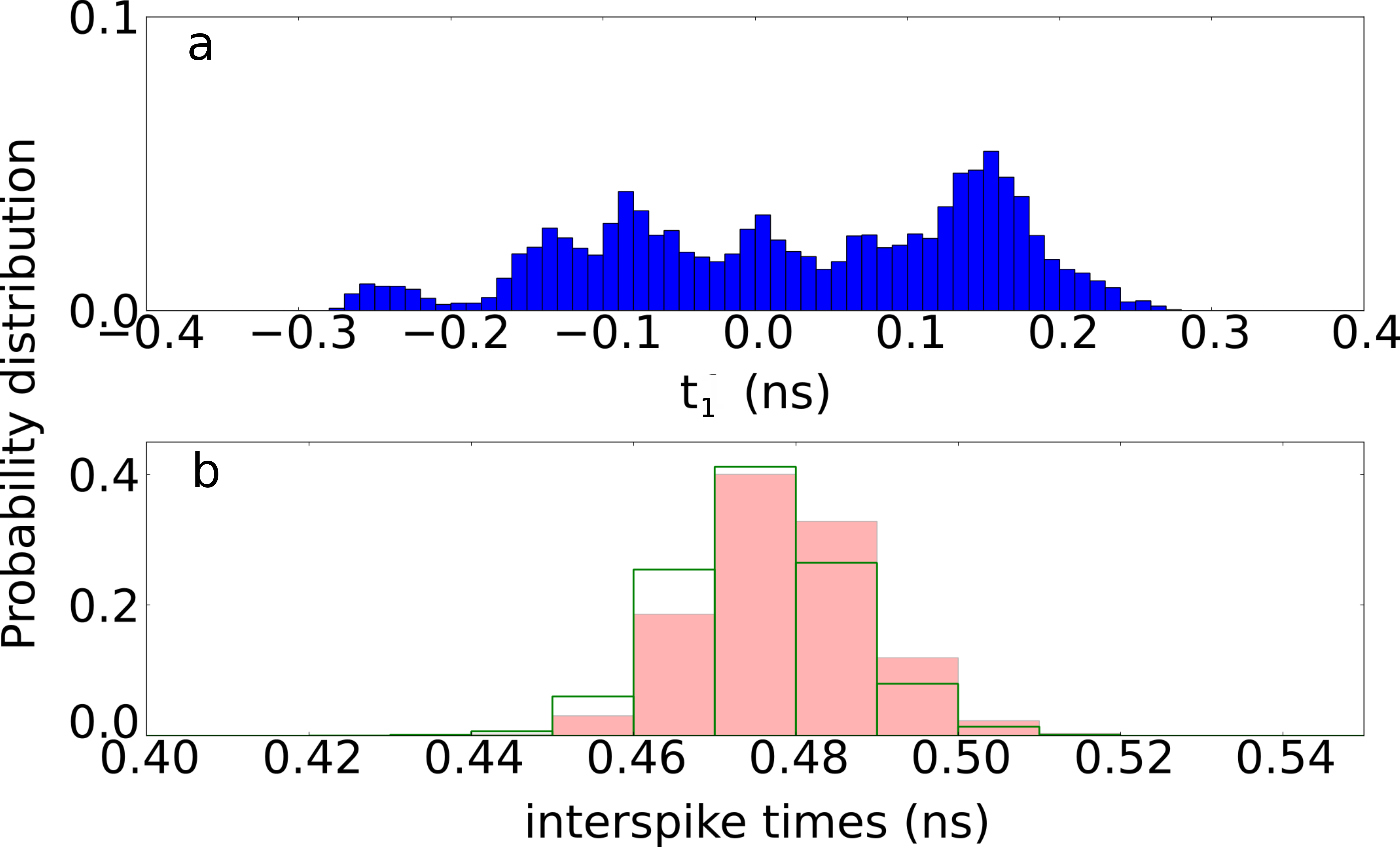}\includegraphics[clip,width=0.49\textwidth]{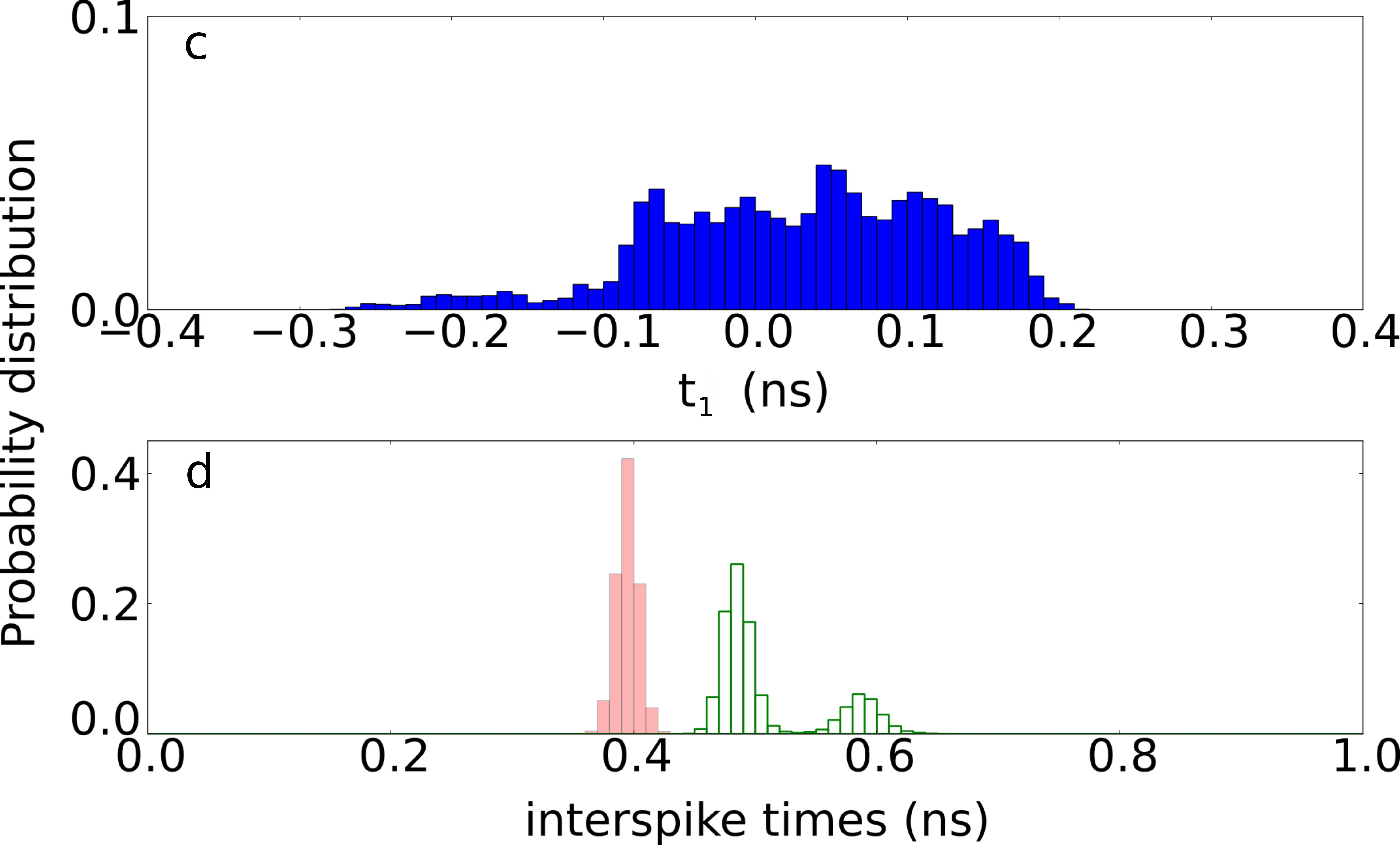}\caption{Panel a: Probability distribution of the position of the first DS
within the cavity ($\Delta_{n}$) for the space-time like diagram
plotted in Fig.~\ref{fig:indep}b. Panel b: Probability distribution
function for the separation between the second and the first DS (pink
trace) and between the third and the second DS (green trace) of the
molecule of Fig.~\ref{fig:indep}b. Panel c: Same as in a) but for
the first DS of the space-time like diagram plotted in Fig.~\ref{molecules}a.
Panel d: Probability distribution function for the separation between
the second and the first DS (pink trace) and between the third and
the second DS (green trace) of the 6-DS molecule of Fig.~\ref{molecules}a.\label{histo} }
\end{figure*}

In the parameter region where vectorial DS exist multiple situations
with different number of independent DS and/or DS molecules evolving
within the external cavity are observed for the same values of the
parameters. This multiplicity is a consequence of the multi-stability
between each DS and the CW solution. In Fig.~\ref{fig:indep}a-c
we have shown three of those situations for $\tau_{f}=2.9\,$ns and
$\tau_{r}=3.3\,$ns. In Fig.~\ref{molecules}a, we depict for the
same parameters a situation where six DS are bound together to form
a molecule of six elements. The molecule exhibits a random walking
as a function of the round-trip covered, wandering inside the cavity,
but the distance between its elements is maintained. In Fig.~\ref{histo},
we analyze statistically the evolution of the 3-DS molecule of Fig.~\ref{fig:indep}b
and the 6-DS molecule shown in Fig.~\ref{molecules}a, as well as
their bond distances. Panels a and c of Fig.~\ref{histo} display
the probability distribution of the position of the first peak of
respectively the 3-DS molecule and of the 6-DS molecule. This curve
has been obtained analyzing the evolution of the molecules on $36\times10^{3}$
round-trips. Both panels indicate that the molecules wander inside
the external cavity in a range of approximately $400\,$ps ($300\,$ps
for the 6-DS molecules) and, in this visited range, the probabilities
of the positions are quite uniform, thus indicating that there are
no preferential positions. The probability distribution of the bond
distances (panel b) for the 3-DS molecule and panel d) for the 6-DS
molecule reveals that the separation between solitons in a molecule
is either maintained during the evolution or jumps between well defined
values. In the case of the 3-DS molecule the bond distance is centered
at 480 ps both for the first to the second soliton and for the second
to the third. This bond distance corresponds to $\tau_{r}-\tau_{f}$,
as explained by the theoretical model. Fluctuations of this bond distance
during the molecule evolution are at the limit of the sampling resolution
of the scope (the standard deviation is approximately 10 ps). 

In the case of the 6-DS molecule the bond distances are peaked to
different values depending on the couple of peaks considered. The
existence of discrete bond distances is a common feature for dissipative
solitons and it was found already in the case of spatial DS \cite{SFA-PRL-00}.
Three values are observed considering all the neighbor distances:
$380\,$ps, $500\,$ps and $600\,$ps (for clarity in panel d) only
the distances between two couples of elements are considered). For
some couple of elements (for example the second and the third peak),
the bond distance can change during the evolution and, in this case,
it may jump to another value. Jumps observed are of $120\,$ps, and
the possible bond distances observed correspond to one of the three
values described before. It is worth remembering that $120\,$ps is
the inverse of the birefringence of the laser used. This suggests
that, while the main bond distances are fixed by $\Delta\tau=\tau_{r}-\tau_{f}$
, multiples, discrete sub-distances exist and are ruled by birefringence.
This feature has been also found in our theoretical analysis.

When the size of the external cavity is increased, larger molecules
and a larger number of independent DS can be hosted in the external
cavity. As examples, in Fig.~\ref{molecules}b,c, we show two situations
obtained for a value of $\tau_{f}=10.8\,$ns and $\tau_{r}=11.1\,$ns.
In panel b) we show a 7-DS molecule which occupies a limited size
of the cavity. The bond distance is given by $2\Delta\tau=600\,$ps.
In panel c) we show three independent DS: two 2-DS molecules, both
having a bond distance of 600 ps, and a single DS.

\subsection{Floquet analysis \label{sub:Floquet-analysis}}

Eqs.~(\ref{eq:X}-\ref{eq:D}) have different periodic solutions,
and their linear stability can be determined via a Floquet analysis,
which implies the reconstruction of the monodromy operator $\mathcal{M}$
for the perturbations to the periodic solutions. A periodic solution
is stable if the whole spectrum of the associated Floquet multipliers
is composed of complex numbers having modulus less than one, and unstable
otherwise. A Floquet multiplier equal to one corresponds to a neutral
mode, and it is typically associated with an invariance of the system,
e.g., translational invariance of a solution. If a periodic solution
consists of $N$ pulses that can be freely displaced one relatively
to the other within the period, $N$ Floquet multipliers equal to
1 must exist. In the case of $N$ independent DS, small deviations
from unity for these multipliers may appear due to the residual interactions
for finite values of the time delay $\tau_{f}$.

In our case, the operator $\mathcal{M}$, although a priori infinite
dimensional, reduces to a square matrix of size $\tau/\delta t$ due
to the discrete sampling incurred by the constant step-size numerical
algorithm. It is constructed by taking one arbitrary point of the
periodic orbit and inserting a small perturbation in all the degrees
of freedom as represented by the mesh points in the delay time and
let the system evolve over an entire period. The deviation of the
end point from the unperturbed orbit yields a column of the operator
$\mathcal{M}$. The eigenvalues (i.e. the Floquet's multipliers) and
eigenvectors of $\mathcal{M}$ can be calculated with a complete decomposition
method (i.e. the QR method \cite{NR-BOOK}), although this is not
efficient for large sizes of $\mathcal{M}$. In our case, $\dim\left(\mathcal{M}\right)\sim10^{5}$
or higher, and it is more efficient to use the Implicitly Restarted
Arnoldi Method \cite{IRAM}. Although this latter method provides
only the dominant eigenvalues (i. e., those with the largest modulus),
it allows us to assess the stability of the periodic solutions and
to compute the dominant Floquet multipliers. We verified that both
methods give identical results. 

The results were found also to be in agreement with the one given
by DDEBIFTool \cite{DDEBT} for the orbits with one and two DS. However,
it was not possible to use such software in general since we do not
have a bifurcation scenario for the apparition of the DS. As such,
we needed to jump-start the Newton process of convergence to the periodic
solution from a guess orbit found via direct numerical integration.
Consequently, we faced the problem of the weak convergence of the
Newton method in a high dimensional space. While correcting orbits
with one or two DS was possible, this process became increasingly
difficult for larger numbers of DS. This is due to the fact that in
each case the adaptive mesh discretization had to be increased to
accommodate for the larger number of localized temporal structures
increasing the dimensionality of the problem.

\subsection{Vectorial phase model\label{sub:Vectorial-phase-model}}

We mention briefly the derivation of the phase model that can be found
in \cite{JMG-PRA-14}. Far from threshold, the fluctuations of the
total intensity die out rapidly and the dynamics remains confined
on a Stokes sphere of a given radius. In addition, strongly elliptical
states would incur a large energetic penalty due to the imbalance
between the two carrier reservoirs. This further confines the residual
dynamics to the vicinity of the equatorial plane of the Stokes sphere.
Without external perturbation, one may not expect any complex residual
dynamics in such situations from Eq.~(\ref{eq:X}-\ref{eq:D}) since
the center manifold is only two dimensional. Besides, the center manifold
consists in two decoupled variables, the polarization orientation
angle $\Phi$ and the optical phase.

Notwithstanding, the coherent delayed retro-actions imposed by the
feedback terms in Eqs.~(\ref{eq:X},\ref{eq:Y}) make it so that
the optical phase of the field couples back into the dynamics. As
such our reduced model will consist in a ``vectorial'' phase for
the orientation of the quasi-linear polarization as well as for the
optical phase of the field. After applying a multiple time scale analysis
to Eq.~(\ref{eq:X}-\ref{eq:D}) considering the relaxation oscillation
as a smallness parameter (see \cite{JMG-PRA-14} for more details),
we obtain a phase model for the orientation, i.e. the difference between
the two circular components of the field $\psi_{\pm}$, $\Phi=\psi_{+}-\psi_{-}$
and the global optical phase represented conveniently by the half
sum $\Sigma=\left(\psi_{+}+\psi_{-}\right)/2$. The reduced model
with PSF and XPR, which is valid only far from threshold, reads

\begin{widetext}
\begin{eqnarray}
\dot{\Sigma} & = & \left(\alpha\gamma_{a}-\gamma_{p}\right)\cos\Phi-\bar{\eta}\sin\frac{\Phi^{\tau_{f}}}{2}\sin\frac{\Phi}{2}\sin\left(u+\Omega+\Sigma-\Sigma^{\tau_{f}}\right)-\bar{\beta}\cos\frac{\Phi}{2}\sin\frac{\Phi^{\tau_{r}}}{2}\sin\left(u+a+\Sigma-\Sigma^{\tau_{r}}\right)\label{eq:Sig}\\
\frac{\dot{\Phi}}{2} & = & \left(\gamma_{a}+\alpha\gamma_{p}\right)\sin\Phi+\bar{\eta}\sin\frac{\Phi^{\tau_{f}}}{2}\cos\frac{\Phi}{2}\cos\left(u+\Omega+\Sigma-\Sigma^{\tau_{f}}\right)-\bar{\beta}\sin\frac{\Phi}{2}\sin\frac{\Phi^{\tau_{r}}}{2}\cos\left(u+a+\Sigma-\Sigma^{\tau_{r}}\right)\label{eq:Phi}
\end{eqnarray}

\end{widetext}

with $\left(\bar{\eta},\bar{\beta}\right)=\left(\bar{\eta},\bar{\beta}\right)\sqrt{1+\alpha^{2}}$
and $u=\text{\ensuremath{\arctan}}\,\alpha$. It is worthwhile to
notice that these two phases $\Phi$ and $\Sigma$ are of very different
nature: while the optical phase precise value $\Sigma$ is irrelevant
due to the phase invariance in an autonomous system the orientation
phase $\Phi$ value is a meaningful quantity. This is due to the broken
rotational invariance imposed by the dichroism and the birefringence
of the VCSEL cavity. 

In the case of a monochromatic solution, the orientation of the quasi
linear polarization $\Phi$ reaches a fix point while the half sum
$\Sigma\sim\omega t$ drifts at the frequency of the mode under consideration.
The modal structure of Eqs.~(\ref{eq:Sig},\ref{eq:Phi}) is a complex
problem and is reported in \cite{JMG-PRA-14}. In the case of the
DS presented in this manuscript, hundreds of such monochromatic solutions
defined by doublets $\left[\omega,\Phi\left(\omega\right)\right]$
exist. The analysis of such modes is useful for explaining the DS
in the sense that we identified that some modes verify the resonance
condition $u+\Omega+\omega\tau_{f}=n\pi$ and $u+a+\omega\tau_{r}=m\pi$.
In this case, by writing $\Sigma=\omega t+\delta$ with $\delta\ll1$,
one can decouple the fluctuations of the optical phase $\delta$ dynamics
from the orientation dynamics and reduce the dynamics to a single
equation for $\Phi$.

\subsection{Multiple binding distances and fast dissipative solitons orbits\label{sub:Fast-DS-and_mol}}

\begin{figure*}[t]
\begin{centering}
\includegraphics[bb=30bp 30bp 650bp 710bp,clip,width=0.3\textwidth]{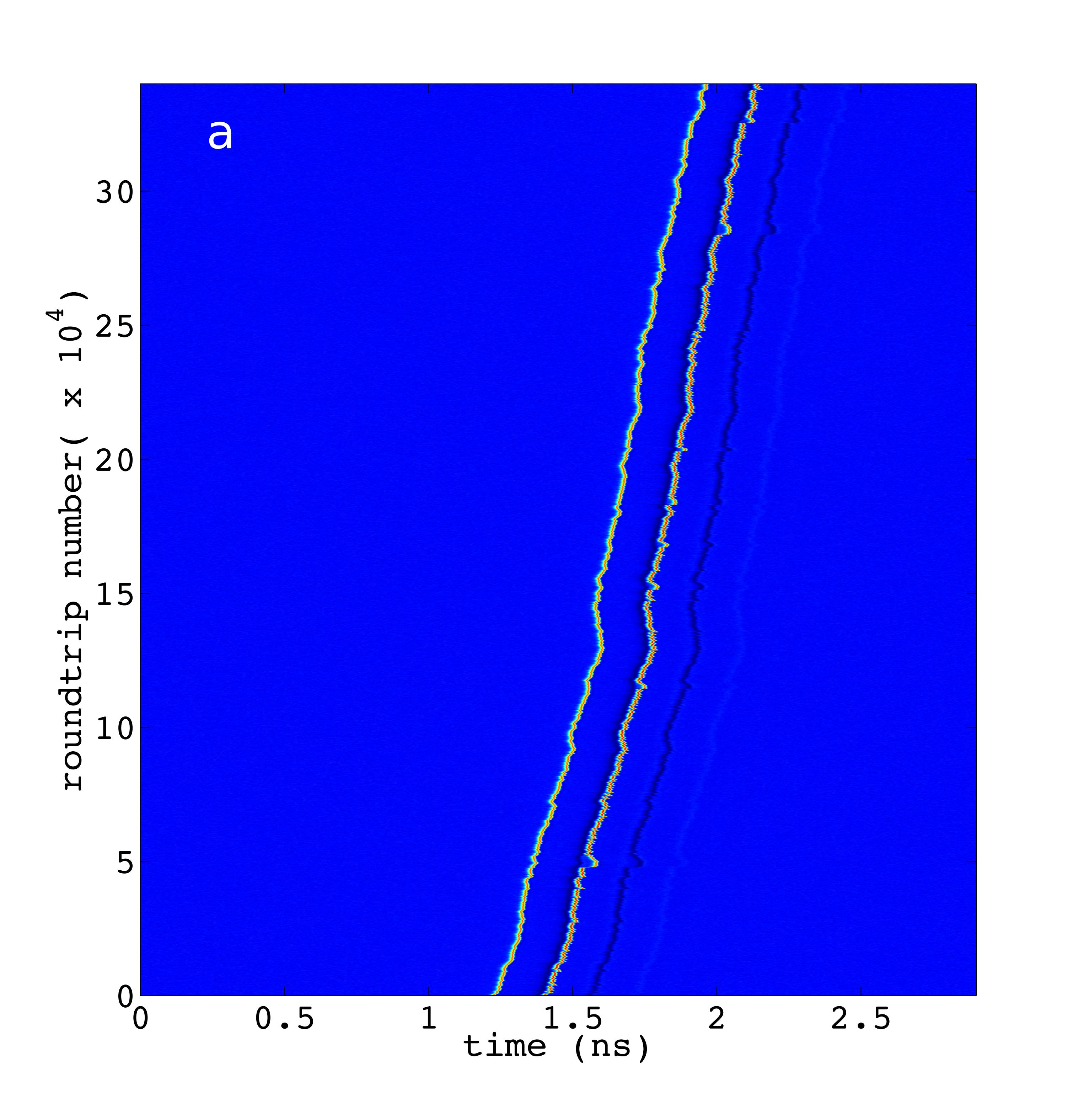}\includegraphics[bb=30bp 30bp 650bp 710bp,clip,width=0.3\textwidth]{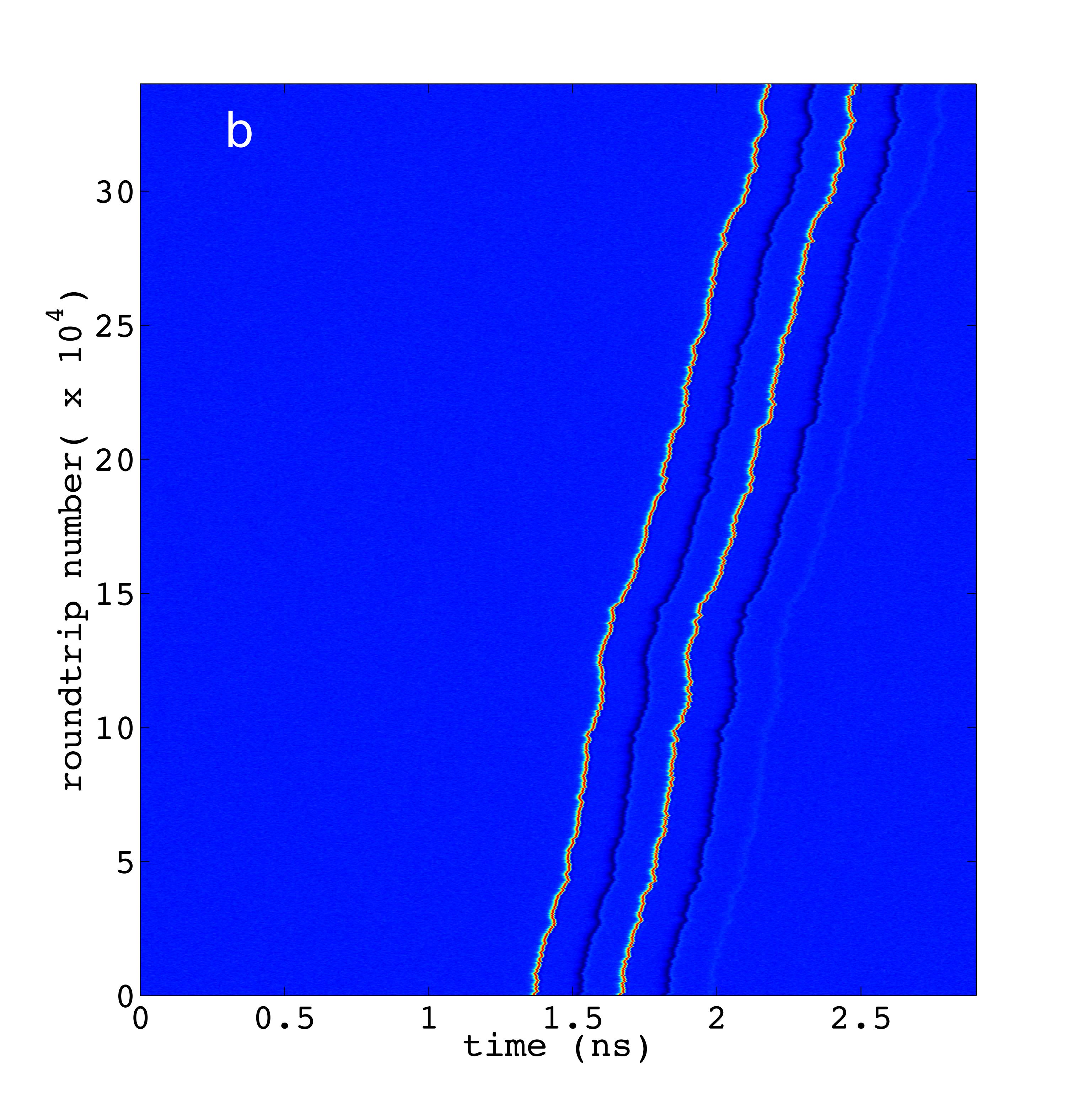}\includegraphics[bb=30bp 30bp 650bp 710bp,clip,width=0.3\textwidth]{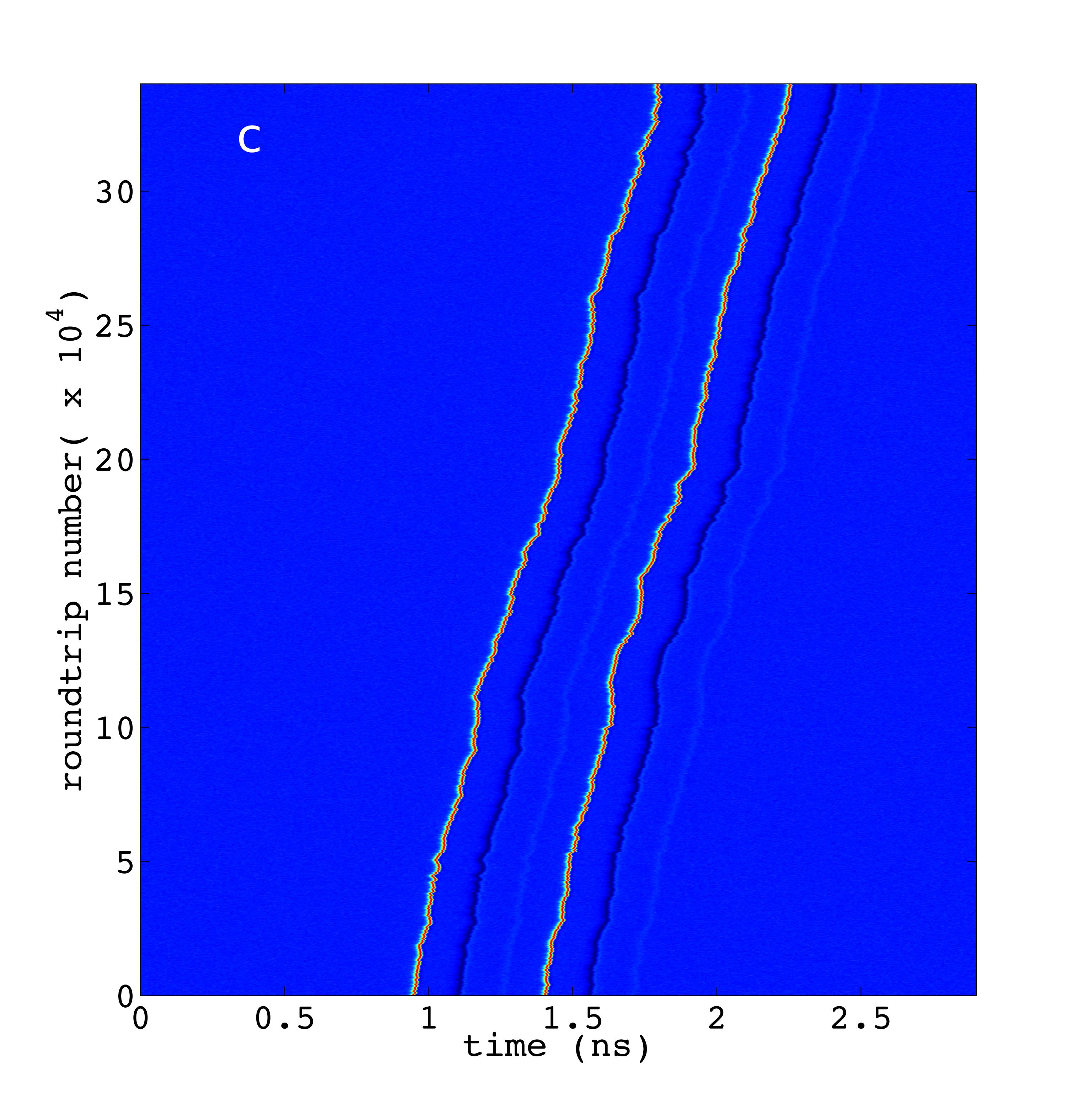}
\par\end{centering}

\begin{centering}
\includegraphics[width=0.9\textwidth]{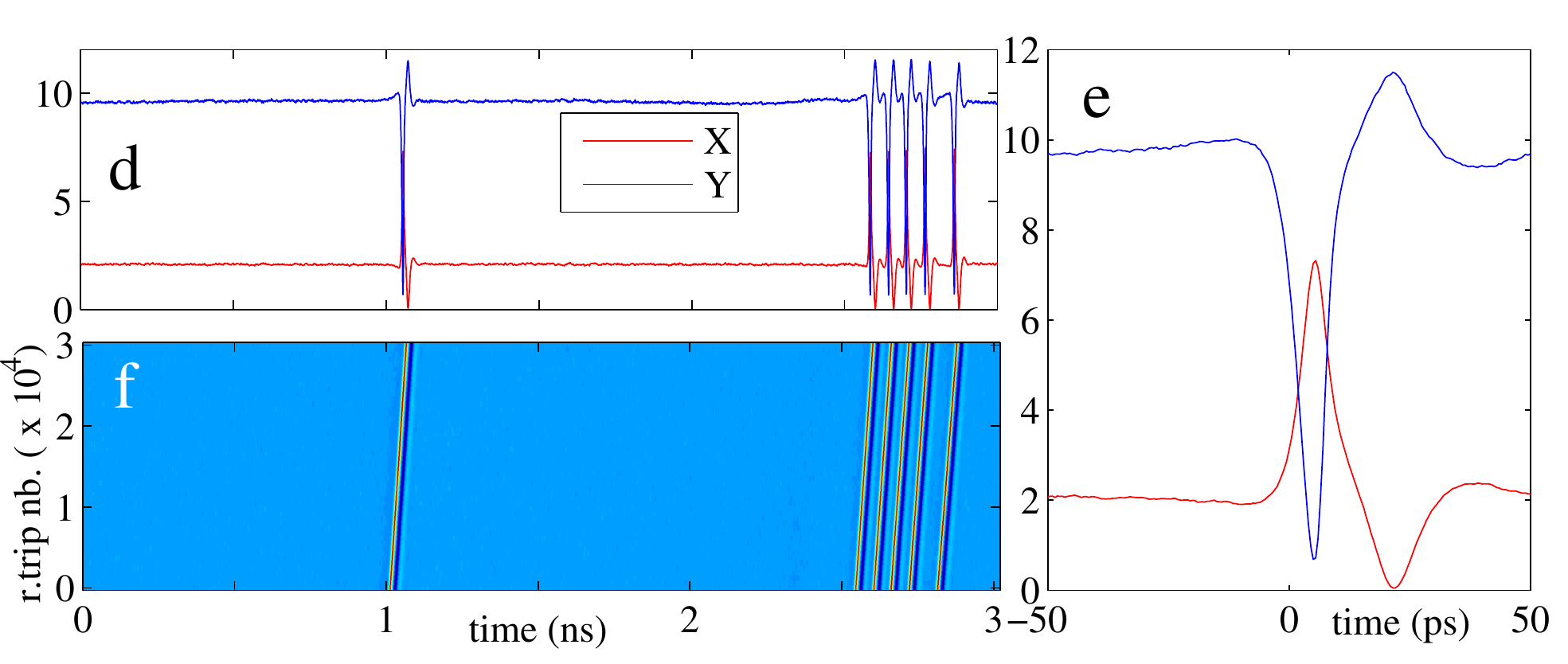}
\par\end{centering}

\centering{}\caption{Several stable equilibrium distances are found and correspond to multiples
of $\Delta\tau$ as represented in panels a,b, and c. The larger the
distance, the weaker the binding force as indicated by the larger
amount of fluctuations between the two elements of the molecule. For
small values of $\Delta\tau$, the DS temporal extent can be reduced
as depicted in the time trace in d) with $\gamma_{p}=0.1$ (corresponding
to a frequency splitting of $15\,$GHz), PSF and XPR rates of $\eta=0.4$
and $\beta=0.26$, and time delays $\tau_{f}=3\,$ns and $\tau_{r}=\tau_{f}+18\,$ps.
The full orbit considering the oscillatory tail represented in d)
is around $25\,$ps. A complex bit sequence is represented in f) over
many round-trips signaling the stability of the setup to store information
as polarization encoded bit sequences. \label{theo_fast} }
\end{figure*}

From the analysis presented in Figs.~\ref{theo1_tracemol},\ref{theo3_Stokes}
one understands that the DS is composed of two plateaus for the polarization
orientation. The length of the second plateau is controlled by the
difference between the two delays $\Delta\tau$. One can appreciate
in Fig.~\ref{theo_fast}a-c that because the replica of the main
anti-phase dip is re-injected after a time $\Delta\tau$, the same
is also true for this replica that is re-injected at time $2\Delta\tau$,
etc. This effect induces discrete binding distances that allows for
the existence of molecules of different types. We present in in Fig.~\ref{theo_fast}a-c,
molecules with binding distance $\Delta\tau$, $2\Delta\tau$ and
$4\Delta\tau$. Obviously, the larger the distance, the smaller the
replica and the weaker the binding force.

However, with the perspective of applications in mind, such multiple
binding distances can be a hindrance and this second plateau can be
reduced as much as the time needed to perform the first part of the
orbit. Here, the secondary kink would be located at a very close distance,
thereby negating this multiplicity of binding distances. This allows
to close the polarization loop in Fig.~\ref{theo3_Stokes}d-f by
performing a single, uninterrupted, orbit. For larger values of the
birefringence and of the PSF and XPR rates we obtain the results described
in Fig.~\ref{theo_fast}d-f, which correspond to DS whose duration
is $\sim25\,$ps. Here only considering the FWHM would yield a much
shorter width of $10\,$ps although such short pulse-width should
not be understood as an inverse effective bit rate. We estimate such
bit-rate to be between $20$ and $40\,$GHz. For instance, we store
in Fig.~\ref{theo_fast}d-f a bit pattern composed of several DS
in order to demonstrate the robustness of the dynamics even in the
presence of noise. Because the dynamics consists in a pure anti-phase,
the carrier lifetime cannot be identified as the time-scale limiting
the pulse-width. In addition, since the phase dynamics proceeds along
the equator of the Stokes sphere or, equivalently, because the polarization
remains always linear, there is no imbalance between the two population
reservoirs indicating that the spin-flip time scale could also not
be the ultimate limiting factor. Finally, because the PSF and the
XPR rates govern the temporal extent of the DS, we think that superior
results could even be achieved by using a VCSOA in an external double
cavity instead of using a VCSEL.
\end{document}